\documentclass[prc,showpacs,showkeys,superscriptaddress,nofootinbib,floatfix,onecolumn]{revtex4}
\usepackage[T2A]{fontenc}
\usepackage{color}
\usepackage{amsfonts}
\usepackage{amsbsy}
\usepackage{mathrsfs}
\usepackage{graphicx}
\usepackage{subfigure}
\newcommand{\eeq}{\end{equation}}
\newcommand{\bea}{\begin{eqnarray}}
\newcommand{\eea}{\end{eqnarray}}

\def\lsim{\mathrel{\rlap{
\lower4pt\hbox{\hskip-3pt$\sim$}}
    \raise1pt\hbox{$<$}}}     %less than approx. symbol
\def\gsim{\mathrel{\rlap{
\lower4pt\hbox{\hskip-3pt$\sim$}}
    \raise1pt\hbox{$>$}}}     %greater than or approx. symbol

\begin{document}

%\DOIsuffix{theDOIsuffix}

%\Volume{XX}
%\Month{XX}
%\Year{XXXX}

%\pagespan{1}{}

%\Receiveddate{XXXX}
%\Reviseddate{XXXX}
%\Accepteddate{XXXX}
%\Dateposted{XXXX}

\title{

Wigner dynamics of quantum semi-relativistic oscillator }

\author{ю.S.~Larkin}
\author{V.S.~Filinov}
\thanks{Corresponding author\quad E-mail:~\textsf{vladimir\_filinov@mail.ru}}
\affiliation{Joint Institute for High Temperatures, Russian Academy of
Sciences, Moscow, Russia}
%\author{Yu.B. Ivanov}%{e-mail: Y.Ivanov@gsi.de}
%\affiliation{GSI Helmholtzzentrum
% f\"ur Schwerionenforschung, % GmbH,
%Planckstr. 1, D-64291
%Darmstadt, Germany}
%\affiliation{Kurchatov Institute,
%Kurchatov sq. 1,
%Moscow, Russia}
%\author[]{Y.B.~Ivanov\inst{3,4}}
%\author[]{V.V.~Skokov\inst{3,5}}
%\author{M. Bonitz}
%\affiliation{Institute for Theoretical Physics and Astrophysics, Christian %Albrechts University, % Kiel,
% Leibnizstrasse 15, D-24098
%Kiel, Germany}
%\author{V.E. Fortov}
%\affiliation{Joint Institute for High Temperatures, Russian Academy of
%Sciences, Moscow, Russia}
%\author{P.R. Levashov}
%\affiliation{Joint Institute for High Temperatures, Russian Academy of
%Sciences, Moscow, Russia}
%
%---------------------------


\begin{abstract}
The integral Wigner - Liouwille equation describing time evolution of the
semi-relativistic quantum 1D harmonic oscillator have been exactly solved
by combination of the Monte-Carlo procedure and molecular dynamics methods.
The strong influence of the relativistic effects on the time evolution of the momentum,
velocity and coordinate Wigner distribution functions and the average values of
quantum operators have been studied. Unexpected 'protuberances'  in
time evolution of the distribution functions were observed.
Relativistic proper time dilation for oscillator have been calculated.
\end{abstract}


\maketitle

\section{Introduction}\label{s:intro}

The harmonic oscillator concept occupies a central position in
science and engineering due to its simplicity and exact solubility
in both classical and quantum descriptions. Today it appears in
mechanics, electromagnetism, electronics, optics, acoustics,
astronomy, nuclear theory and so on. In the quantum approach the
harmonic oscillator is one of the exactly solvable problems studied
in detail due to its considerable physical interest and
applicability. The problem of a one-dimensional harmonic oscillator
 is also one of the important paradigms leading to the enormous
applications in a wide range of the modern physics \cite{HO}.

 There
are at least three logically autonomous alternative paths to
quantization. The first is the standard one utilizing operators in
Hilbert space, developed by Heisenberg, Schrodinger, Dirac, and
others in the 1920s. The second one relies on path integrals, and
was conceived by Dirac and constructed by Feynman \cite{feynm}. The
third one is the phase-space formulation. It is based on Wigner▓s
(1932) quasi-distribution function and Weyl▓s (1927) correspondence
between ordinary c-number functions in phase space and quantum
mechanical operators in Hilbert space \cite{tatr1}. This complete
formulation is based on the Wigner function (WF), which is a
quasiprobability distribution function in phase-space.

Wigner▓s quasi-probability distribution function in phase-space is a special (Weyl-
Wigner) representation of the density matrix. It has been useful in describing transport properties,
quantum optics, nuclear physics, quantum computing, decoherence and chaos.
%It is
%also of importance in signal processing, and the mathematics of algebraic deformation.
It furnishes a third, alternative, formulation of quantum mechanics,
independent of the conventional Hilbert space, or path integral formulations.
In this logically complete and self-standing formulation, one need not choose sides between
coordinate or momentum space. It works in full phase-space, accommodating the
uncertainty principle; and it offers unique insights into the classical limit of quantum theory:
The variables (observables) in this formulation are c-number functions in phase space
instead of operators, with the same interpretation as their classical counterparts, but are
composed together in novel algebraic ways.

The simplest and perhaps most straightforward generalization of nonrelativistic
quantum theory towards the inclusion of relativistic kinematics leads
to Hamiltonians $\hat{H}$ that involve the relativistic kinetic energy, or relativistically covariant form
of the free energy, of a particle of mass $m$ and momentum $p$, given by the square-root operator
and a coordinate-dependent static interaction potential $V (x)$, so
 $%\begin{eqnarray} \label{RHam}
\hat{H}(\hat{p},\hat{x}) = \sqrt{\hat{p}^2c^2 + m^2c^4} + V(\hat{x}.)
 $ ($c$ is velocity of light in free space).
The eigenvalue equation of this Hamiltonian is usually called the ⌠spinless Salpeter equation.■
It may be regarded as a well-defined approximation to the Bethe√Salpeter formalism \cite{SB1} for the
description of bound states within relativistic quantum field theories, obtained when assuming
that all bound-state constituents interact instantaneously and propagate like free particles \cite{SB2}.
Among others, it yields semi-relativistic descriptions of hadrons as bound states of quarks
\cite{LS1,LS2}.

We study rigorously the semi-relativistic quantum 1D harmonic oscillator described by
the Hamiltonian operator
composed of the relativistic kinetic energy and a static
harmonic potential using Wigner
formulation of quantum mechanics. In this research we have
considered a time evolution of such system, notably the time evolution
of the momentum, coordinate and velocity distributions, average values of their
quantum operators and relativistic time dilation. We used both
Monte-Carlo procedure and method of molecular dynamic for numerical
solution of this problem.

%----------------------------

\section{Wigner - Lioville equation} \label{WL}
{\bf Integral form of the Wigner - Lioville equation}.
The most conventional formulation of quantum mechanics is
description of system's dynamics with complex wave function. This
function of particle with mass $m$ and moving in potential field
$V(x)$ (we consider one dimension) satisfies the Schroedinger
equation:
\begin{eqnarray}\label{Shred}
i\hbar\frac{\partial\Psi(q,t)}{\partial t} = \hat H\Psi(q,t)
\end{eqnarray}
,where hamiltonian is $\hat H = \hat p^2/2m + V(\hat q)$ and the initial
condition is $\Psi(q,0)=\Psi_0(q)$

More general description of quantum systems is given in terms of the
density matrix
$\rho(q,q';t)$ (in coordinate representation), which has the following
form in case of pure state: $\rho(q,q';t) = \Psi(q,t)\Psi^*(q',t)$.
Evolution equation for density matrix is
\begin{eqnarray}\label{Shred1}
i\hbar\frac{\partial\rho(q,q';t)}{\partial t} = (\hat H - \hat
H'^*)\rho(q,q';t), \label{Shred1}
\end{eqnarray}
where $\hat H$ acts on coordinate $q$, while $\hat H'$ acts on
coordinate $q'$. The initial condition for this equation has form
$\rho(q,q';0) = \rho_0(q,q')=\Psi_0(q)\Psi_0^*(q')$.

In the Wigner representation of quantum mechanics we use joint
distribution of quasiprobability for momentum and coordinate; it is
called Wigner function \cite{tatr1}. Wigner function is
defined as Fourier transform of density matrix on difference
variable $\xi=q-q'$, while center varible is $x=(q+q')/2$:
\begin{eqnarray}\label{WL}
W(x,p;t) = \frac{1}{2\pi\hbar}\int\limits_{-\infty}^{\infty}{\rho(x
- \xi/2,x + \xi/2)e^{ip\xi/\hbar}} d{\xi}.
\end{eqnarray}

Evolution of quantum system in Wigner representation is describing
by Wigner - Liouwille equation \cite{tatr1}:

\begin{eqnarray} \label{WLE}
\frac{\partial W(x,p;t)}{\partial t} + \frac{\partial
H(p,x)}{\partial p}\frac{\partial W(x,p;t)}{\partial x} -
 \frac{\partial H(p,x)}{\partial x}\frac{\partial W(x,p;t)}{\partial p} =
 \nonumber\\
 \frac{1}{i\hbar}\left[ V\left(x + \frac{i\hbar}{2}\frac{\partial}{\partial p}\right) -
            V\left(x - \frac{i\hbar}{2}\frac{\partial}{\partial
            p}\right) - i\hbar V'_x(x)\frac{\partial}{\partial p}
            \right]W(x,p;t),
\end{eqnarray}
with initial condition $W(x,p;0) = W_0(x,p)$ defined by  (\ref{WL}) at $t=0$.
When the related to Hamiltonian the classical
Hamilton's function has the form $H(p,q) = p^2/2m + V(x)$, the
partial derivatives are: $\partial H/\partial x = V'_x(x)$,
$\partial H/\partial p = p/m $.

In quantum mechanics Wigner functions are real valued but altering sign analog of the
probabilistic joint $p$ and $x$ distributions in classical mechanics.
This is supported by its general properties
%of Wigner function
\cite{tatr1}: a) density of probability in momentum space is
$%\begin{eqnarray}\label{WWL}
W(p) = \int\limits_{-\infty}^{\infty}{}\,W(x,p)dx
$;  %\end{eqnarray}
 b) density of probability in configuration space is
$%\begin{eqnarray}\label{WWL1}
W(x) = \int\limits_{-\infty}^{\infty}{}\,W(x,p)dp \quad $.
%\end{eqnarray}
In additition $W(x,p)$ is bilinear in wave function $\Psi(x,t)$ or in $\Psi(p,t)$ in
momentum representation;

One can rewrite evolution equation (\ref{WLE}) in the integral form
\cite{tatr1,kmf}:
\begin{eqnarray}\label{INTFW1}&&
W(x,p;t) = \int G(p,x,t;p_0,x_0,0)W_0(x_0,p_0)\,dp_0 \,dx_0 +
\nonumber\\&& \int_0^td\tau \int dp_{\tau} dx_{\tau} G
\left(p,x,t;p_{\tau},x_{\tau},\tau\right) \int_{-\infty}^\infty
ds\,W\left(x_{\tau}, p_{\tau} - s, \tau \right) \omega
\left(s,x_{\tau} \right)
%\frac{1}{2\pi}\int\limits_{0}^{t}\,dt' \int \,dp'\,dx'
%\int\,d\lambda\,dp_0\lambda G(p,x,t;p_0,x',t')\times
%\nonumber\\
%\sin{\lambda(p_0-p')}\left[ V'_x(x') - \frac{1}{\lambda\hbar}\left(
%V\left( x' + \frac{\lambda\hbar}{2} \right) - V\left( x' -
%\frac{\lambda\hbar}{2} \right) \right) \right]W(x',p',t').
\end{eqnarray}
%\begin{eqnarray}\label{INTFW1}
%W(x,p;t) = \int G(p,x,t;p_0,x_0,0)W_0(x_0,p_0)\,dp_0 \,dx_0 +
%\frac{1}{2\pi}\int\limits_{0}^{t}\,dt' \int \,dp'\,dx'
%\int\,d\lambda\,dp_0\lambda G(p,x,t;p_0,x',t')\times
%\nonumber\\
%\sin{\lambda(p_0-p')}\left[ V'_x(x') - \frac{1}{\lambda\hbar}\left(
%V\left( x' + \frac{\lambda\hbar}{2} \right) - V\left( x' -
%\frac{\lambda\hbar}{2} \right) \right) \right]W(x',p',t').
%\end{eqnarray}
with $\omega \left(s,x \right)= F\left(x\right) \frac{d\delta
\left(s\right) }{ds} + \frac {4}{\left(2\pi \right)}\int d{q} \,
V\left(x-q\right) \sin \left(2sq\right)$. Here $F\left(x\right) = -
\partial V(x)/\partial x$  is the classical force.
%, while $V$ is potential energy.
Here $G(p,x,t; p_\tau,x_\tau,\tau)$ is the Green's function for classical
Liouwille equation
\begin{eqnarray}\label{FGR}
G(p,x,t; p_\tau,x_\tau,\tau) = \delta(p - \bar p(t;\tau,p_\tau,x_\tau))\delta(x -
\bar x(t;\tau,p_\tau,x_\tau)),
\end{eqnarray}
where $\bar p$,  $\bar x$ are solutions of Hamilton's equations
\begin{eqnarray}\label{HEQ1}
\frac{d p}{dt} = -\frac{\partial H(p,x)}{\partial x}, \qquad
\frac{d x}{dt} = \frac{\partial H(p,x)}{\partial p},
\end{eqnarray}
with initial conditions
$%\begin{eqnarray}\label{INHEQ}
\bar p(\tau;\tau,p_\tau,x_\tau)=p_\tau, \bar x(\tau;\tau,p_\tau,x_\tau)=x_\tau,
$.

Solution of equation (\ref{INTFW1}) can be written in the form of the
iterative series, which have the following interpretation. The first term of
the series (the first term in the r.h.s of the Eqs~\ref{INTFW1}) is equal to
the sum of the contributions of the virtual classical trajectories defined
by the dynamical Hamilton's equations (\ref{HEQ1}). The contribution of the
each virtual trajectory is equal to the value of the initial Wigner function
$W_0(x_0,p_0)$ taken at initial point $x_0,p_0$. Next terms of the iterative
series are equal to the sum of
the contributions of virtual trajectories consisting of
segments of classical trajectories, separated by 'jumps' in momentum.
The term's number in iterative series is equal to number of the momentum
'jumps' as it follows from convolution structure of integral term in
equation (\ref{INTFW1}). In the classical limit ($\hbar\to 0$) the force term in
$\omega$ cancels the last term and only the first term of iterative series gives
the main contribution to solution of the Wigner-Liouville equation.
In the classical limit Eqs.~(\ref{WLE}), (\ref{INTFW1}) are reduced to the
classical Liouville equations.
%In quantum case this form allows
%to write the Eq.~(\ref{WLE}) as the classical Liouville equation plus a quantum
%term.

%\subsection{Restrictions on initial condition $W_0(x,p)$. }
{\bf Restrictions on initial condition}. To find solutions of the
integral equation (\ref{INTFW1}) the
initial function $W_0(x,p)$ have to be taken according to the
definition (\ref{WL}). However this definition imposes the certain restriction
on the choice of possible functions in phase space. This can be easily illustrated for
harmonic oscillator (particle in potential field $V(x) = V_0 + V_1x + V_2x^2$
). In this case the Wigner-Liouwille equation (\ref{WLE}) is reduced to the form
\begin{eqnarray}\label{HGOS}
\frac{\partial W(x,p;t)}{\partial t} + \frac{p}{m}\frac{\partial
W(x,p;t)}{\partial x} - V'_x(x)\frac{\partial W(x,p;t)}{\partial p}=
0,
\end{eqnarray}
and coincides with the classical Liouwille equation. However
solution of quantum harmonic oscillator is totally distinguished from
classical one. Consequently, quantum and classical solutions of this
task distinguish from each other in choice of the initial condition
$W_0(x,p)$. So we need additional condition to choose classical or
quantum solution. For density matrix of pure state this condition
can be formulated as follows
%In case of pure state "the term of choice of quantum
%solution" can be written as (in terms of density matrix)
\begin{eqnarray}\label{ro}
\frac{\partial^2}{\partial x_1\partial x_2}\ln{\rho(x_1,x_2;0)} =0,
\mbox{ or in Wigner representation \cite{tatr1} } \quad
\frac{\partial^2}{\partial x_1\partial
x_2}ln\int\limits_{-\infty}^{\infty}W_0(\frac{x_1+x_2}{2},
p)e^{ip(x_1-x_2)/\hbar}\,dp.
\end{eqnarray}
%or in Wigner representation \cite{tatr1}:
%\begin{eqnarray}\label{ro1}
%\frac{\partial^2}{\partial x_1\partial
%x_2}ln\int\limits_{-\infty}^{\infty}W_0(\frac{x_1+x_2}{2},
%p)e^{ip(x_1-x_2)/\hbar}\,dp.
%\end{eqnarray}
%This term holds in following moments of time.
In \cite{klim} one can find another condition of the choice $W_0(x,p)$, which is
equivalent to this one. One of the physical interpretation of this additional conditions is connected
with requirement for mean-square coordinate and momentum deviations to satisfy Heisenberg's
principle of uncertainty at time equal to zero. % must be keeped.

%    \subsection{Calculation of  average physical quantities. Properties of Wigner function.}
{\bf Physical quantities}.
To calculate the average value of physical quantity $<\hat A>$ corresponded to quantum operator $\hat A$
the Weil's symbol $A( p, x)$ has to be introduced by expression \cite{tatr1}:
\begin{eqnarray}\label{weyl}
A( p, x) = \int\limits_{-\infty}^{\infty}{\exp{ip\xi/\hbar}
\left\langle x - \frac{\xi}{2} \left| \hat A \right| x +
\frac{\xi}{2} \right\rangle}\,d\xi \quad
\mbox{then}\quad
<\hat A> = \int{W(x,p)A(p,x)}\,dp\,dx,
\end{eqnarray}
%Weil's symbol of quantum operator $\hat A$ has form \cite{tatr1}:
%\begin{eqnarray}\label{weyl}
%A( p, x) = \int\limits_{-\infty}^{\infty}{\exp{ip\xi/\hbar}
%\left\langle x - \frac{\xi}{2} \left| \hat A \right| x +
%\frac{\xi}{2} \right\rangle}\,d\xi.
%\end{eqnarray}
%With function $A( p, x)$, one can calculate an average value of
%physical quantity, which is corresponded to operator $\hat A$:
%\begin{eqnarray}\label{SrA}
%            <\hat A> = \int{W(x,p)A(p,x)}\,dp\,dx,
%\end{eqnarray}
where $W(x,p)$ is Wigner function.
%\newpage

%\subsection {Non-relativistic harmonic oscillator}
{\bf Non-relativistic harmonic oscillator}.
Non-relativistic harmonic oscillator with mass $m$ and circular frequency
 $\omega$ has Hamiltonian
\begin{eqnarray}\label{Harm}
\hat H = \frac{\hat p^2}{2m} + \frac{m\omega^2 \hat x^2}{2};
\end{eqnarray}
Wigner-Liouville equation (\ref{WLE}) in case of harmonic potential
$V(x) =m\omega^2x^2/2$ has a simple form:
\begin{eqnarray}  \label{Oss}
            \frac{\partial W(x,p;t)}{\partial t} +
                  \frac{p}{m}\frac{\partial W(x,p;t)}
                  {\partial x} - m\omega^2x\frac{\partial W(x,p;t)}{\partial p}
                 = 0,
\end{eqnarray}
As the initial condition for equations (\ref{Oss}) we will
consider a coherent state of harmonic oscillator given by the wave function \cite{quantum} :
$%\begin{eqnarray} \label{PsiN}
            \Psi(x,0) = \left( \frac{m\omega}{\pi\hbar}
            \right)^{1/4}\exp{\left\{ \frac{i\tilde p x}{\hbar} - \frac{m\omega(x - \tilde x)^2}{2\hbar} \right\}}
            \exp{\left\{ -\frac{i\tilde p\tilde x}{2\hbar}
            \right\}}.
$ %\end{eqnarray}
 Here $\tilde p$,$\tilde x$ are average values of momentum and
coordinate at the initial moment $t = 0$.
According to definition (\ref{WL}), the initial Wigner function is defined by:
\begin{eqnarray} \label{WLN}
            W_0(x, p) = \frac{1}{\pi\hbar}\exp{\left\{ -\frac{m\omega(x - \tilde x)^2}{\hbar} -
            \frac{(p - \tilde p)^2}{\hbar m\omega} \right\}}.
\end{eqnarray}
Solution of equation  (\ref{Oss}) as well as the related integral equation (\ref{INTFW1})
can be written in the form
\begin{eqnarray} \label{IOss}&&
W(x,p;t) =
            \int %G(p,x,t;p_0,x_0,0)
(\delta(p - \bar p(t;0,p_0,x_0)) \delta(x - \bar x(t;0,p_0,x_0)))
            W_0(x_0,p_0)\,dp_0
            \,dx_0=
  \nonumber\\&&
            \frac{1}{\sqrt{2\pi(\hbar/2m\omega)}}
            \exp{\left\{ -\frac{m\omega}{\hbar}
           [x - (\tilde x\cos{\omega t} + \frac{\tilde p}{m\omega}sin{\omega
           t})]^2-\frac{1}{\hbar m\omega}[p - (\tilde p cos{\omega t} - m\omega\tilde x sin{\omega t})]^2\right\}}.
\end{eqnarray}
where $\bar p(t)$, $\bar x(t)$ are virtual trajectories defined by
the Hamilton's function:
$%\begin{eqnarray} \label{RHam}
            H(p,x) = \frac{ p^2}{2m} + \frac{m\omega^2 x^2}{2}
$ %\end{eqnarray}
and the Hamilton's equations (\ref{HEQ1})
with initial conditions $\bar p(0) = p_0$, $\bar x(0) = x_0$:
$%\begin{eqnarray} \label{OS}
 \,      \bar p(t) = p_0\cos{\omega t} - m\omega
            x_0\sin{\omega t}, % \qquad
 \bar x(t) = \frac{p_0}{m\omega}\sin{\omega t} + x_0\cos{\omega t}.
$%\end{eqnarray}

From (\ref{IOss}) it follows that in a coherent state of
harmonic oscillator the average momentum and coordinate satisfy classical
law of motion
$%\begin{eqnarray} \label{PSR}
            <p(t)> = \tilde p\cos{\omega t} - m\omega\tilde x\sin{\omega
            t}, \qquad
            <x(t)> = \tilde x\cos{\omega t} + \frac{\tilde p}{m\omega}sin{\omega
           t}
$%\end{eqnarray}
 \,  and the standard deviations of momentum and coordinate are constant:
$%\begin{eqnarray} \label{PSRR}
            <\delta p^2> = \frac{\hbar m\omega}{2},  \qquad
            <\delta x^2> = \frac{\hbar}{2m\omega}
$.  %\end{eqnarray}
While Heisenberg's formula of uncertainty has it's minimum:
$%\begin{eqnarray} \label{HSB}
             <\delta p^2> <\delta x^2> = \hbar^2/4
$%\end{eqnarray}
.    Average energy is constant equal to
$%\begin{eqnarray} \label{SrE}
            E = \frac{\hbar\omega}{2} + \frac{\tilde p^2}{2m} +
            \frac{m\omega^2\tilde x^2}{2}
$%\end{eqnarray}
. \, More general case of the composite states slightly distinguished
from pure coherent state is considered in \cite{kmf}.

%\subsection{Semi-relativistic harmonic oscillator}
{\bf Semi-relativistic harmonic oscillator}.
The Hamiltonian of the semi-relativistic harmonic oscillator
has the form:
\begin{eqnarray}
\label{ORHAM} \hat{H}=\sqrt{\hat{p}^2c^2 + m^2c^4} + \frac{m\omega^2
\hat x^2}{2}.
\end{eqnarray}
Wigner function % must satisfy properties $1$ - $3$ and
has to be a solution of the Wigner - Lioville equation (\ref{WLE}) \cite{tab3}:

\begin{eqnarray} \label{RGOWL}
         \frac{\partial W(x,p;t)}{\partial t} + \frac{pc^2}{\sqrt{p^2c^2 + m^2c^4}}\frac{\partial W(x,p;t)}{\partial
         x} - m\omega^2x\frac{\partial W(x,p;t)}{\partial p} = 0,
\end{eqnarray}
with initial condition $W(x,p;0) = W_0(x,p)$ and the Hamilton's function:
$%\begin{eqnarray} \label{RHam}
            H(p,x) = \sqrt{p^2c^2 + m^2c^4} +
            \frac{m\omega^2x^2}{2}.
$ %\end{eqnarray}

As before solution of this equation looks like:
\begin{eqnarray} \label{IRGOWL}
 W(x,p;t) =
            \int G(p,x,t;p_0,x_0,0)W_0(x_0,p_0)\,dp_0
            \,dx_0,
\end{eqnarray}
where in the Green's function the virtual trajectories $\bar p(t;t_0,p_0,x_0)$
and $\bar x(t;t_0,p_0,x_0)$ are solutions of Hamilton's equations (\ref{HEQ1}). %:
When $c\to\infty$ the Hamilton's function %formula (\ref{RHam})
is equal to its non-relativistic limit (with rest energy term
$mc^2$)
$%\begin{eqnarray} \label{RHam1}
            H\to{mc^2 + \frac{p^2}{2m} + \frac{m\omega^2x^2}{2}}
$. %\end{eqnarray}

%Let us note that in relativistic case the virtual trajectories can't
%be obtained analytically in explicit form, as before.
%However one can find  the
Period of oscillations of the virtual trajectories
versus energy have been found in \cite{tabl}. Indeed from (\ref{HEQ1}) one can obtained an equation
for $p(t)$ and it's first integral:
\begin{eqnarray} \label{Rnut}
\frac{d^2\bar{p}}{dt^2} + m\omega^2c\frac{\bar{p}}{\sqrt{\bar{p}^2 + m^2c^2}} = 0, \qquad
            C_1 = \left(\frac{d\bar{p}}{dt}\right)^2 + 2m\omega^2\sqrt{\bar{p}^2c^2 +
            m^2c^4},
\end{eqnarray}
where $C_1 = 2m\omega^2E $, energy $E = \sqrt{p_0^2c^2 + m^2c^4} +
m\omega^2x_0^2/2$ has to be founded from initial conditions $p_0$ and
$x_0$.
 So, after integration of this equation,
the period of oscillations depends on energy of the trajectory
\begin{eqnarray} \label{TRnut}
T(E) = \frac{4\sqrt{2}\sqrt{\frac{E}{mc^2} +
            1}}{\omega} E\left(\frac{E - mc^2}{E + mc^2} \right)
            - \frac{4\sqrt{2}}{\omega\sqrt{\frac{E}{mc^2} + 1}} K\left(\frac{E - mc^2}{E + mc^2}
            \right).
\end{eqnarray}
Here $K(z)$ and $E(z)$ are elliptical integrals of the first and
second kind \cite{ryzhik}:
\begin{eqnarray} \label{TRnut1}
  K(z) = \int_{0}^{\pi/2}\frac{d\phi}{\sqrt{1 -
            z\sin^2{\phi}}},  \qquad
%\nonumber\\
  E(z) = \int_{0}^{\pi/2}{\sqrt{1 -
            z\sin^2{\phi}}}\,d\phi.
\end{eqnarray}
This dependence is represented by the right panel of the Fig.~\ref{imp}.

%\subsection{Method of numerical simulation}
{\bf Numerical simulation}.
We are going to obtain and compare results for quantum non-relativistic and
semi-relativistic harmonic oscillators with the same initial Wigner function,
which corresponds to
a coherent state of non-relativistic harmonic oscillator
(for simplicity we chose %with initial average coordinate $\tilde x_0$ and
the average momentum $\tilde p_0=0$ see (\ref{WLN})):
\begin{eqnarray} \label{RGOWLN}
        W_0(x_0,p_0) = \frac{1}{\pi\hbar}exp{\left[ -\frac{m\omega(x_0 - \tilde
 x_0)^2}{\hbar} - \frac{p_0^2}{\hbar m\omega} \right]}.
\end{eqnarray}

 As we has mentioned before, an evolution of semi-relativistic oscillator is
 described by the expression (\ref{IRGOWL}).
%One can interpret this  by the next way.
To consider the time evolution of oscillator  the following
numerical procedure has  been used. The initial Wigner function
(\ref{RGOWLN}) was considered as
 probabilistic distribution of points ($p_0$,$x_0$) in phase
 space. To sample these points we used Monte Carlo procedure.
Each ($p_0$,$x_0$) point was considered as the initial point of the virtual
dynamic trajectory, described by  Hamilton's equations (\ref{HEQ1}).
To solve these equations
we used molecular dynamics method. Distribution of virtual
trajectories in phase space allow to obtain Wigner function $W(x,p;t)$
at any time $t$.

 Average values of general quantum operators
can be obtained by calculation of the time dependences of the Weyl's symbol
of operators along the virtual trajectories and averaging over ensemble of
all trajectories. For average values of momentum, coordinate, energy
and mean -square values of the momentum and coordinate these averaging have been done
at each time in considered time evolution interval.
%Thereby task of calculation of the time  evolution of any quantum operator for
% semi-relativistic harmonic oscillator with initial state ($\ref{RGOWLN}$) can be
% numerically solved.

In our simulations we have generated $10^5$ of virtual trajectories.
For simulation time dynamics we used implicit finite-difference scheme
with centering \cite{potter} :
    \begin{eqnarray}\label{NumSh}
        p_{m(i+1)} = p_{m(i)} - \frac{\tau}{2}[x_{m(i+1)} +
        x_{m(i)}] , \qquad
        x_{m(i+1)} = x_{m(i)} + z\frac{\tau}{2}{\left[\frac{p_{m(i+1)}}
        {\sqrt{z(z+{p_{m(i+1)}}^2)}} + \frac{p_{m(i)}}
        {\sqrt{z(z+{p_{m(i)}}^2)}}\right]}.
\end{eqnarray}
    Here $\tau = 0.01$ is a time step; this system of algebraic
    equations was mainly solved by  method of simple iterations.
    For numerical calculations we used the following system of unities:
$%\begin{eqnarray}\label{Unit}
        p = \frac{\hbar\omega}{c}\sqrt{z} p_m, \quad
        x = \frac{c}{\omega}\frac{x_m}{\sqrt{z}},  \quad
         t = \frac{c^2}{\hbar\omega} t_m,
$% \end{eqnarray}
where $p_m$, $x_m$, $t_m$ are values in scheme
(\ref{NumSh}) for $\hbar = 1$, $c = 1$
However, all results below are written in usual physical units.

    Physical dimensionless parameter $z = mc^2/\hbar\omega$ defines
    "degree of relativism" \, of the oscillator,
    when $z >> 1$ relativistic effects almost disappear.

%---------------------------------------------------------------------------------------------------------------------------
%---------------------------------------------------------------------------------------------------------------------------
\section {Numerical results}

\begin{figure}[htb]
    %\vspace{0cm} \hspace{0.0cm}
   \includegraphics[width=6.0cm,clip=true]{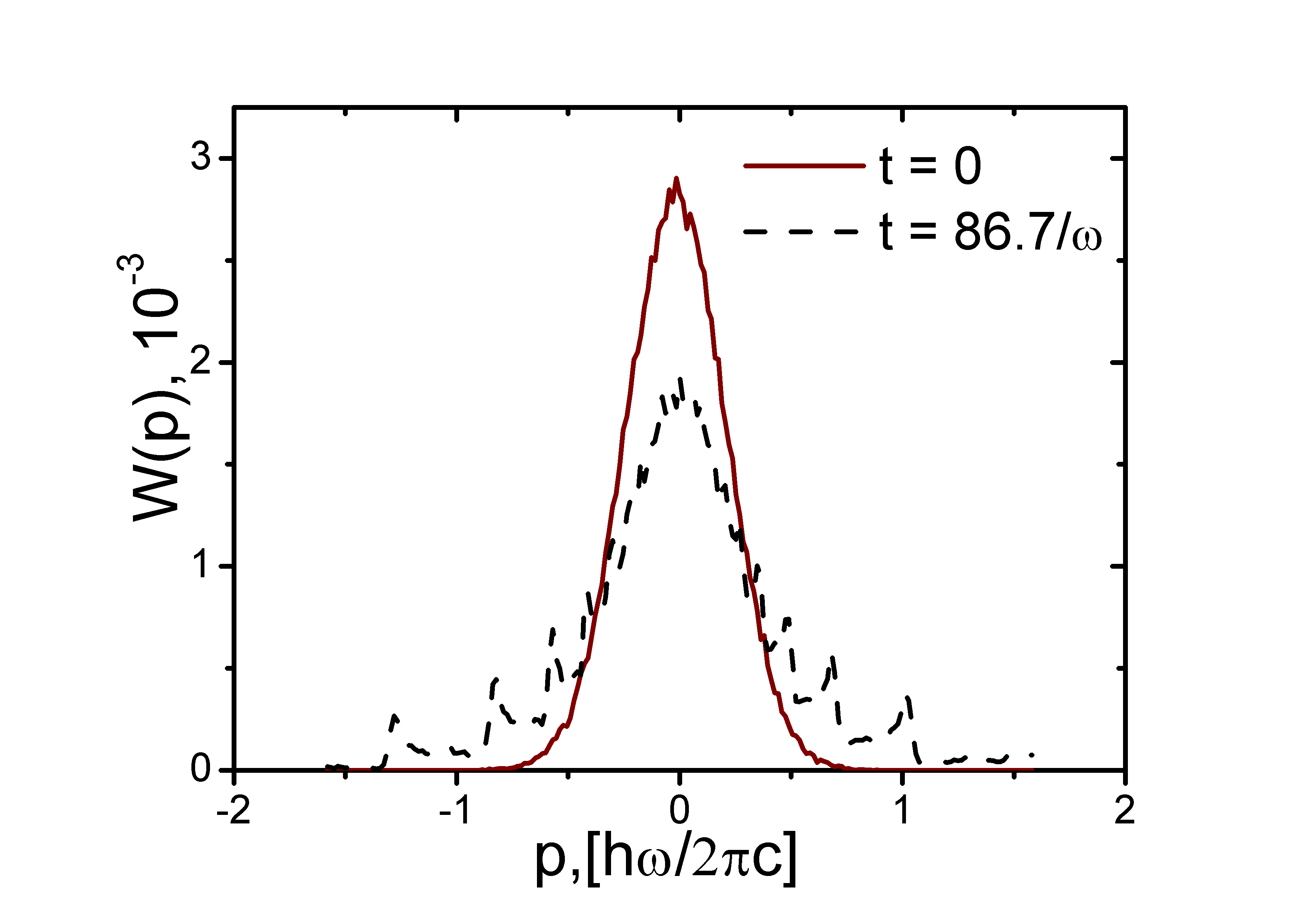}
  \includegraphics[width=6.0cm,clip=true]{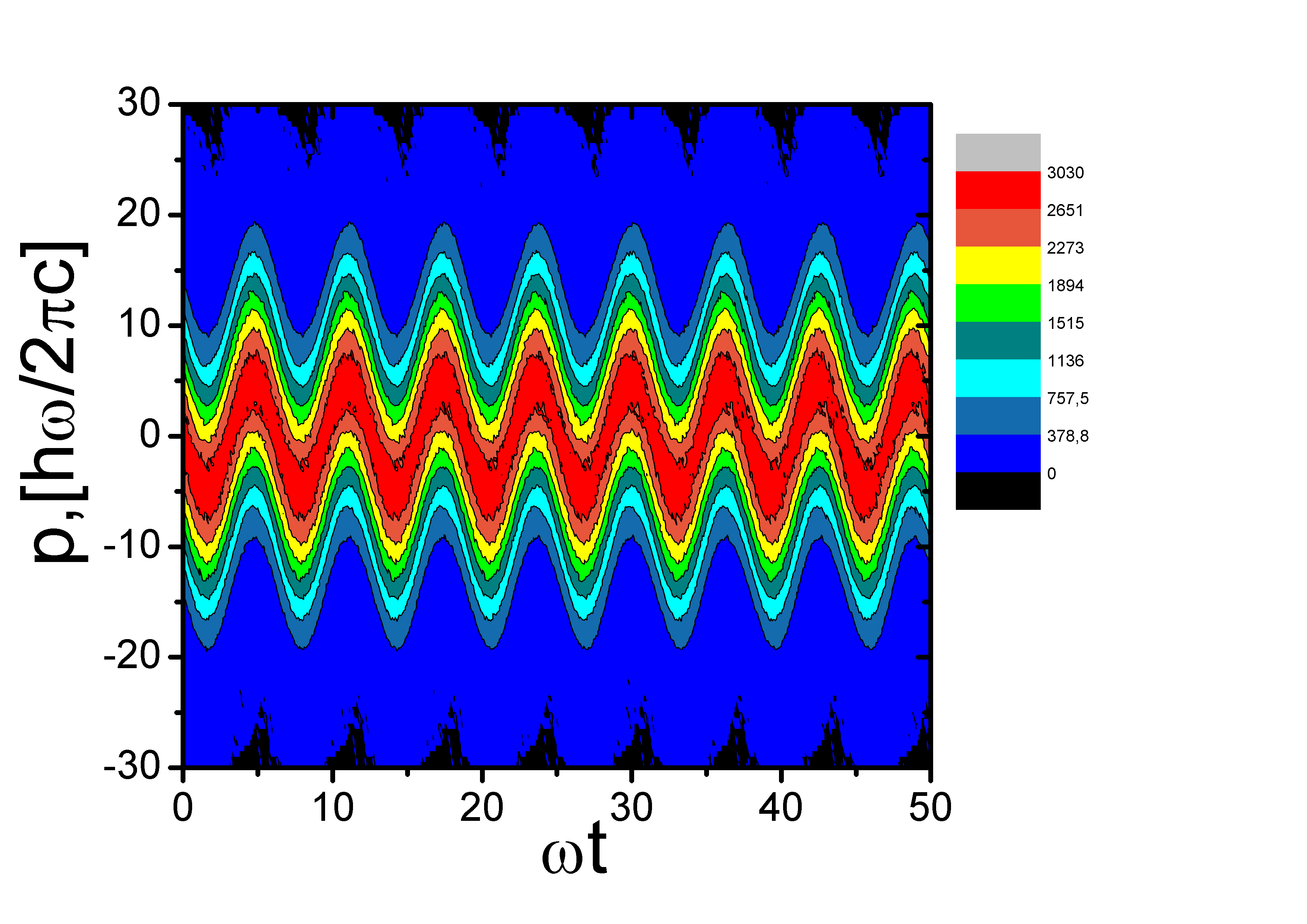}
  \includegraphics[width=6.0cm,clip=true]{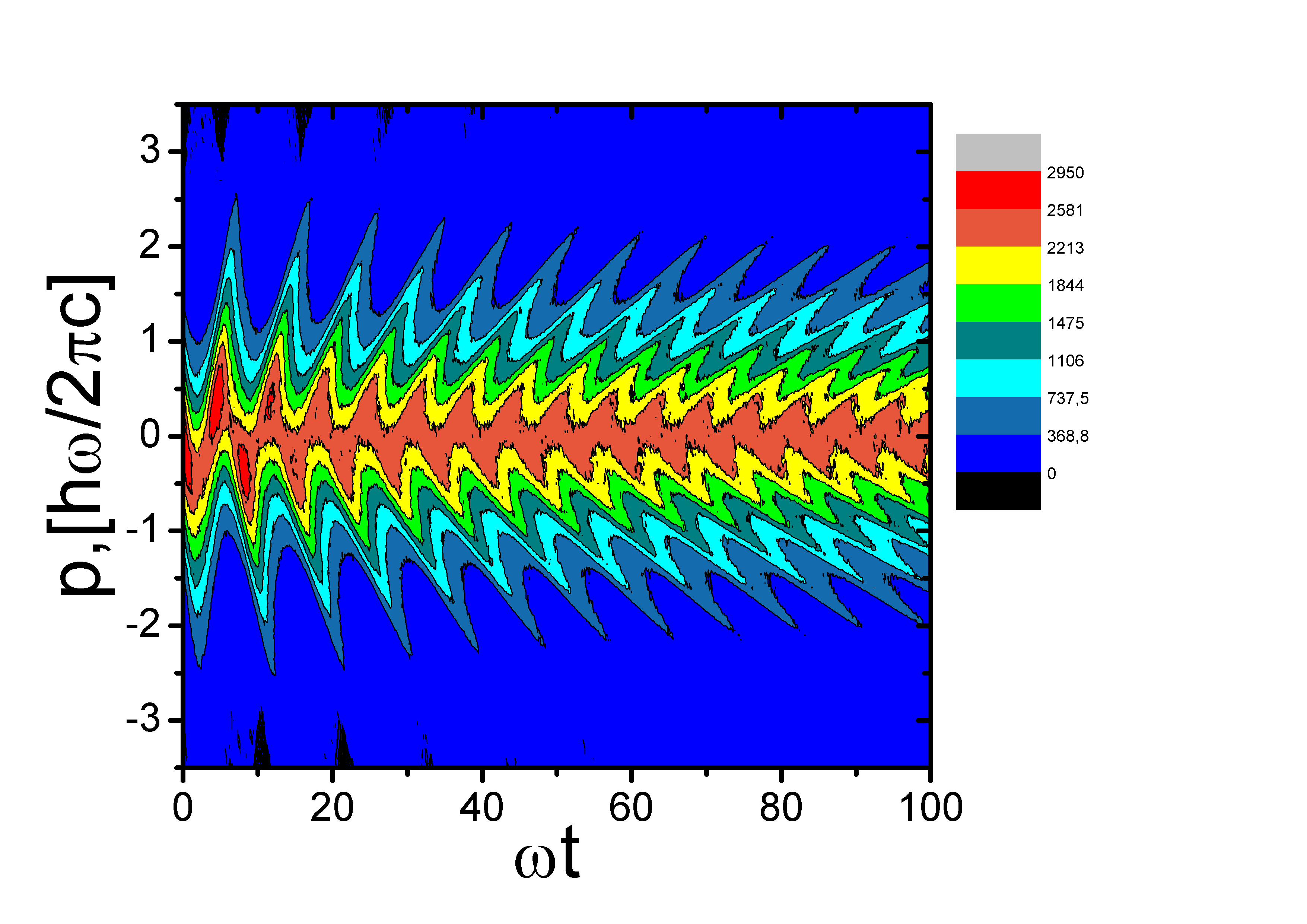}
  \includegraphics[width=6.0cm,clip=true]{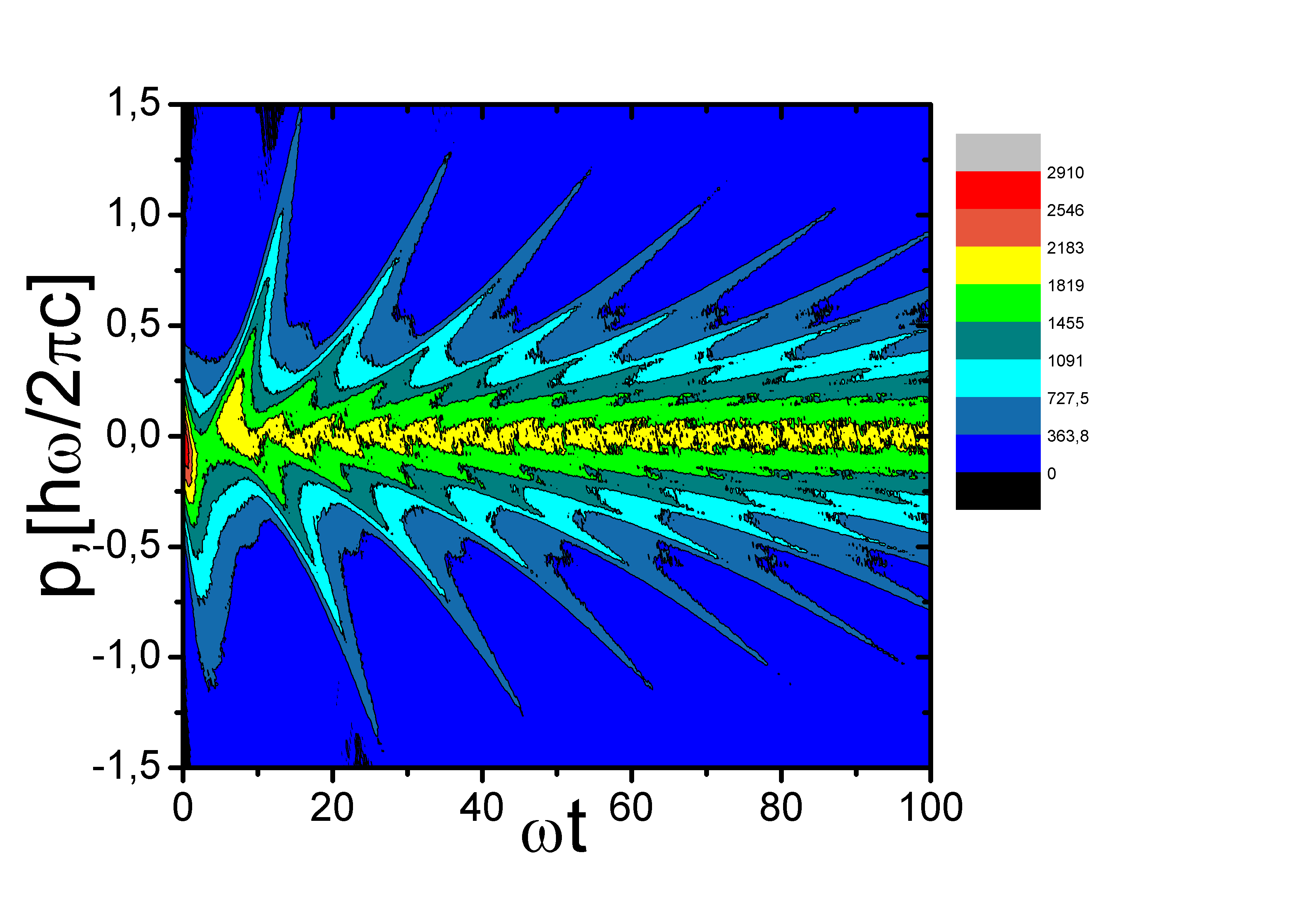}
\caption{Time evolution of the non normalized momentum distributions
for oscillators at $z=100$, $z=1$ and  $z=0.1$. The left top panel
-- initial at $t = 0$ (solid line) and at $\omega t = 86.7$ distributions
(dot line) for $z=0.1$ . The right top panel -- the contour
plot for $z=100$, the left bottom panel -- the contour plot for
$z=1$, the right  bottom panel -- the contour plot for $z=0.1$.
Small oscillations on the initial distribution at $t=0$ characterize
the error of the generated Gauss momentum distribution
(\ref{RGOWLN}). } \label{Imtm}
    \end{figure}

\begin{figure}[htb]
    %\vspace{0cm} \hspace{0.0cm}
    \includegraphics[width=6.0cm,clip=true]{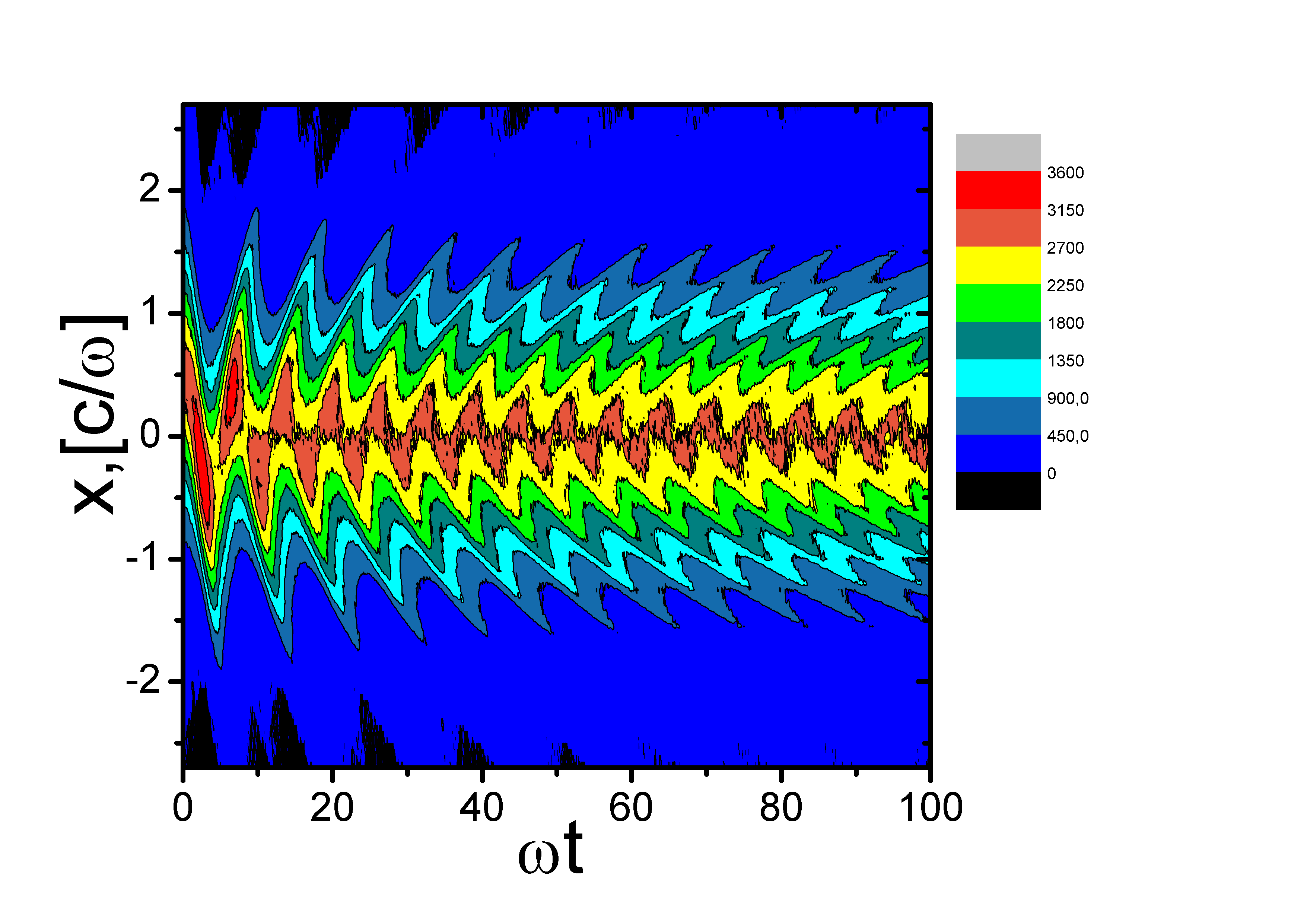}
    \includegraphics[width=6.0cm,clip=true]{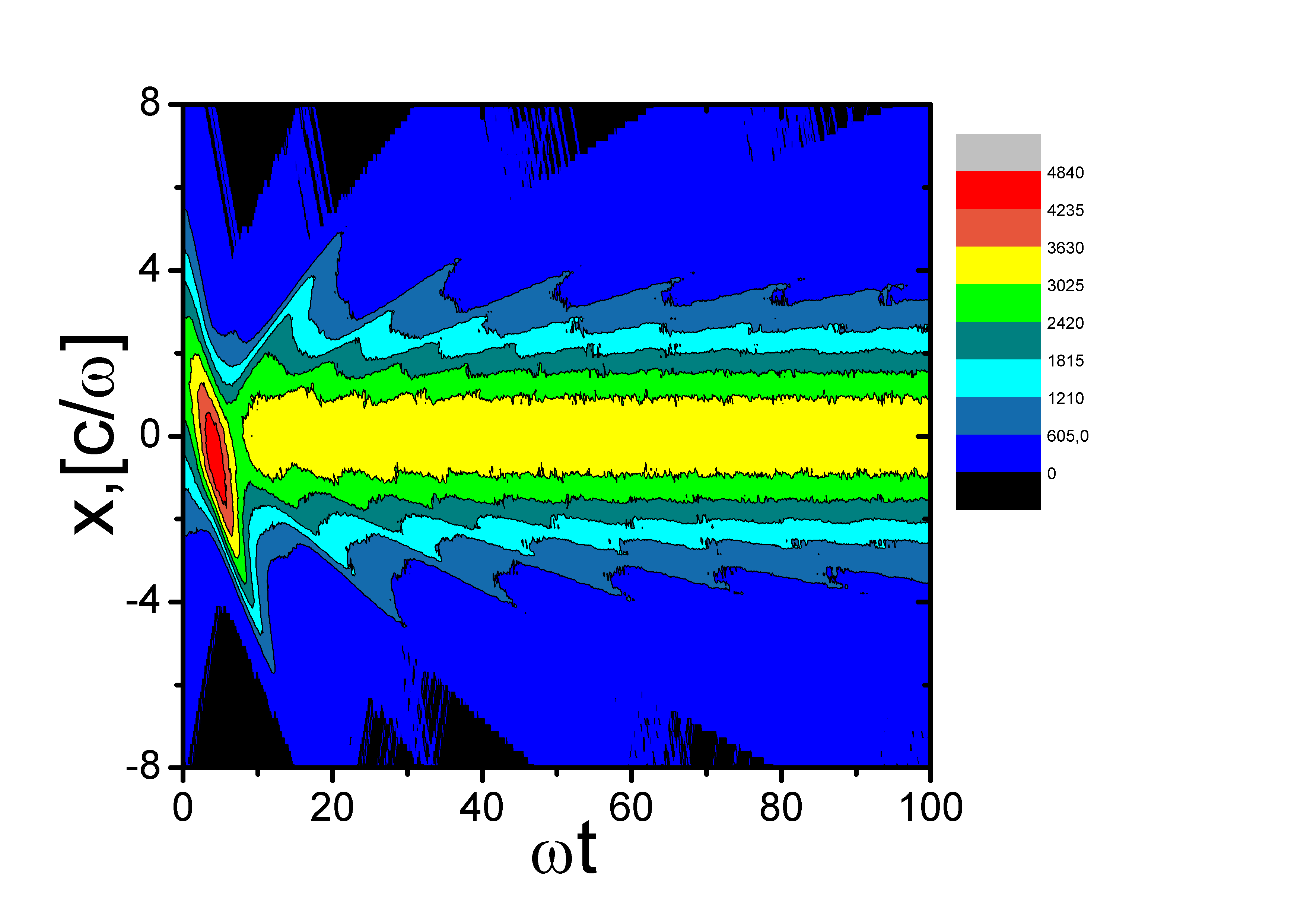}
\caption{The time contour plot of the coordinate non normalized distributions.
Left panel --  $z=1$, right panel -- $z=0.1$.
    }
\label{Sumq}
    \end{figure}

\begin{figure}[htb]
    %\vspace{0cm} \hspace{0.0cm}
    \includegraphics[width=6cm,clip=true]{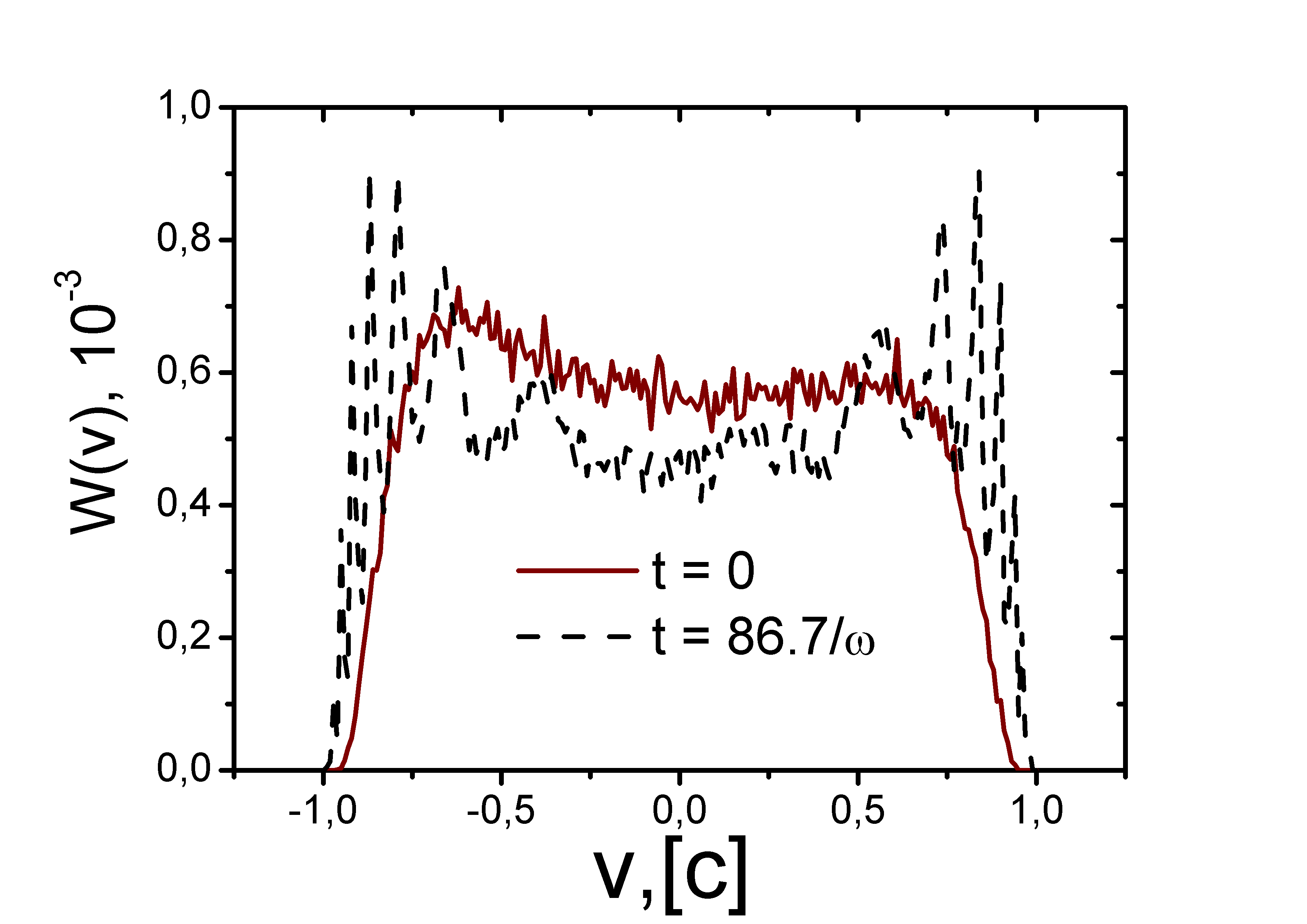}
    \includegraphics[width=6cm,clip=true]{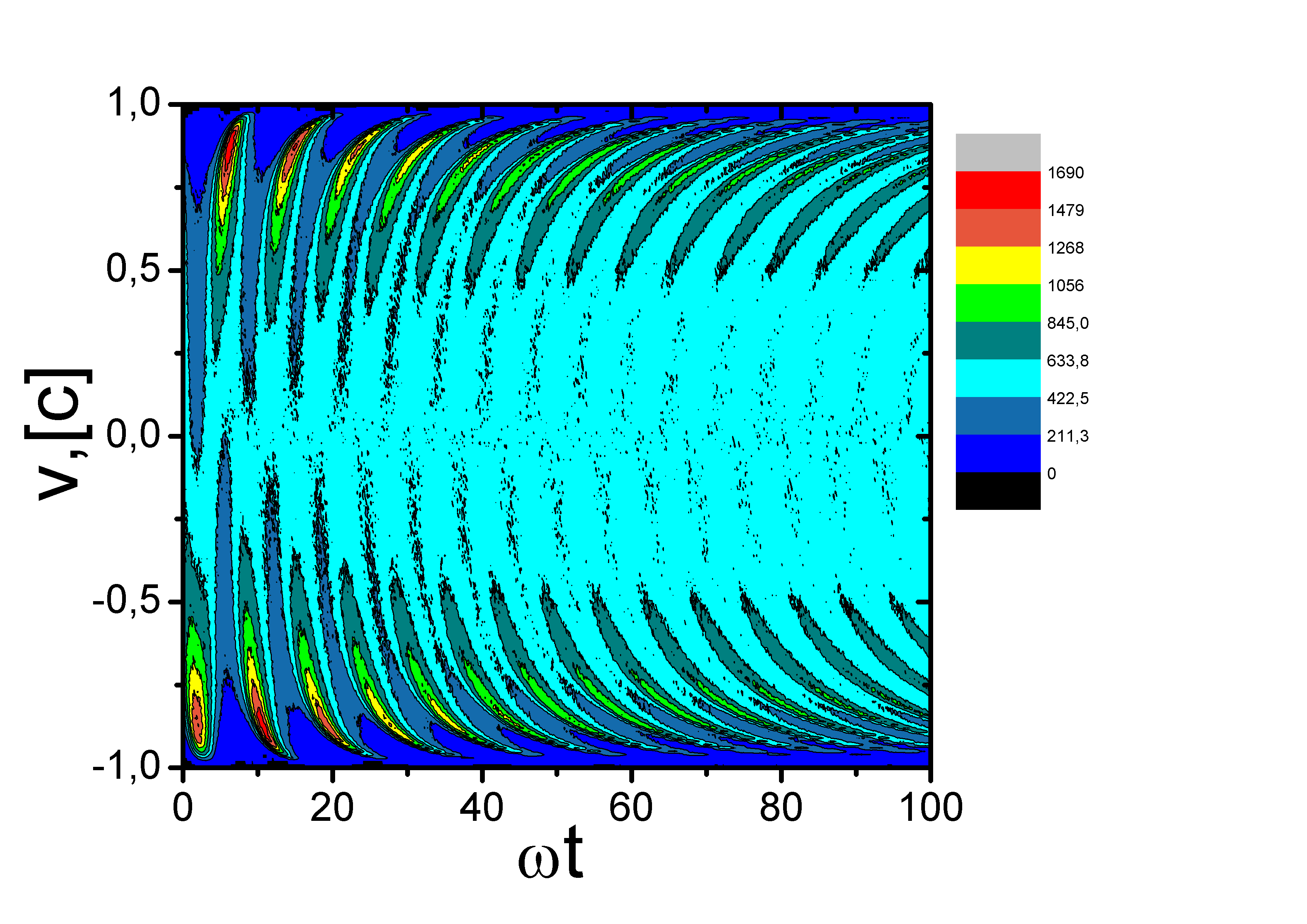}
    \includegraphics[width=6cm,clip=true]{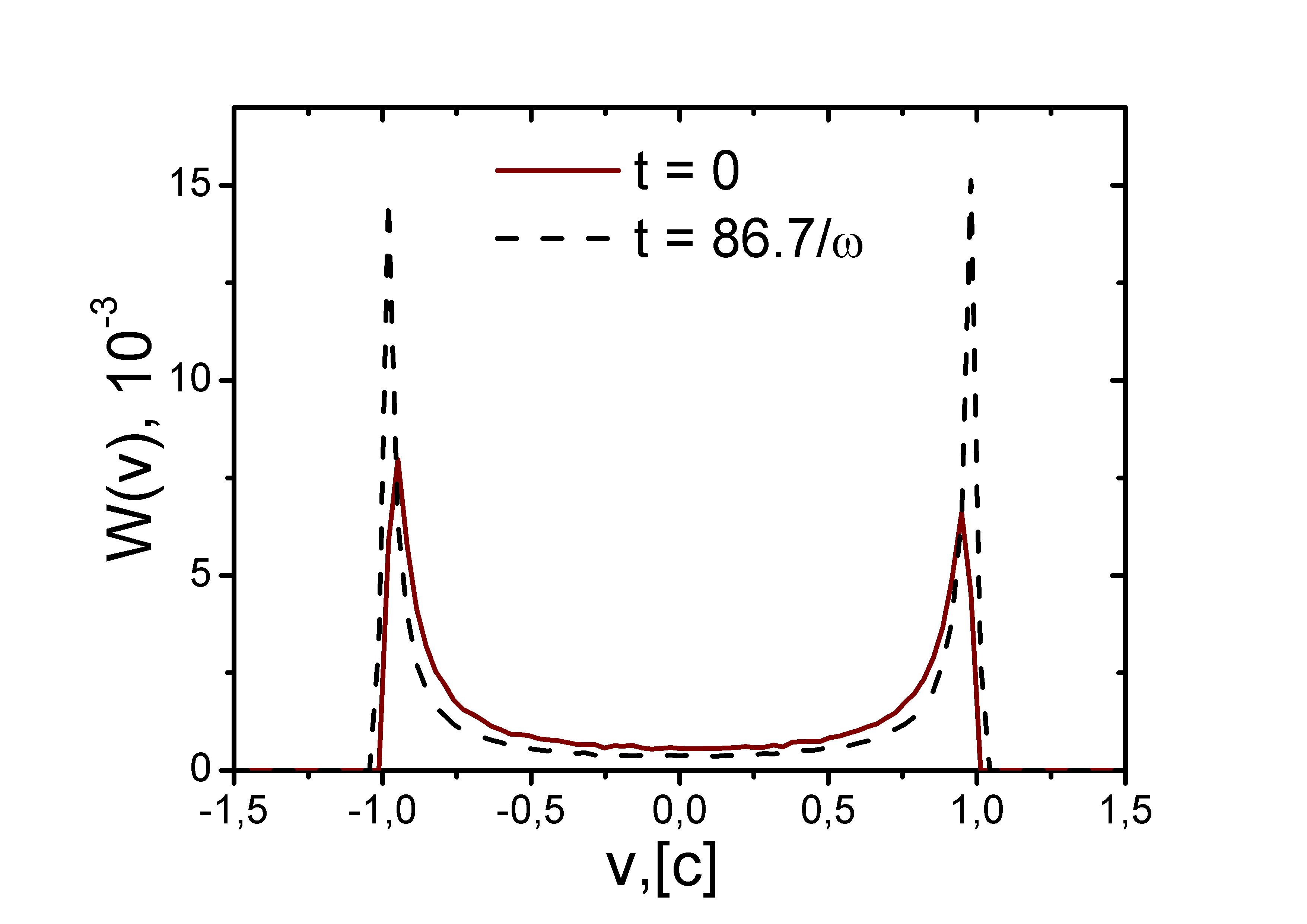}
    \includegraphics[width=6cm,clip=true]{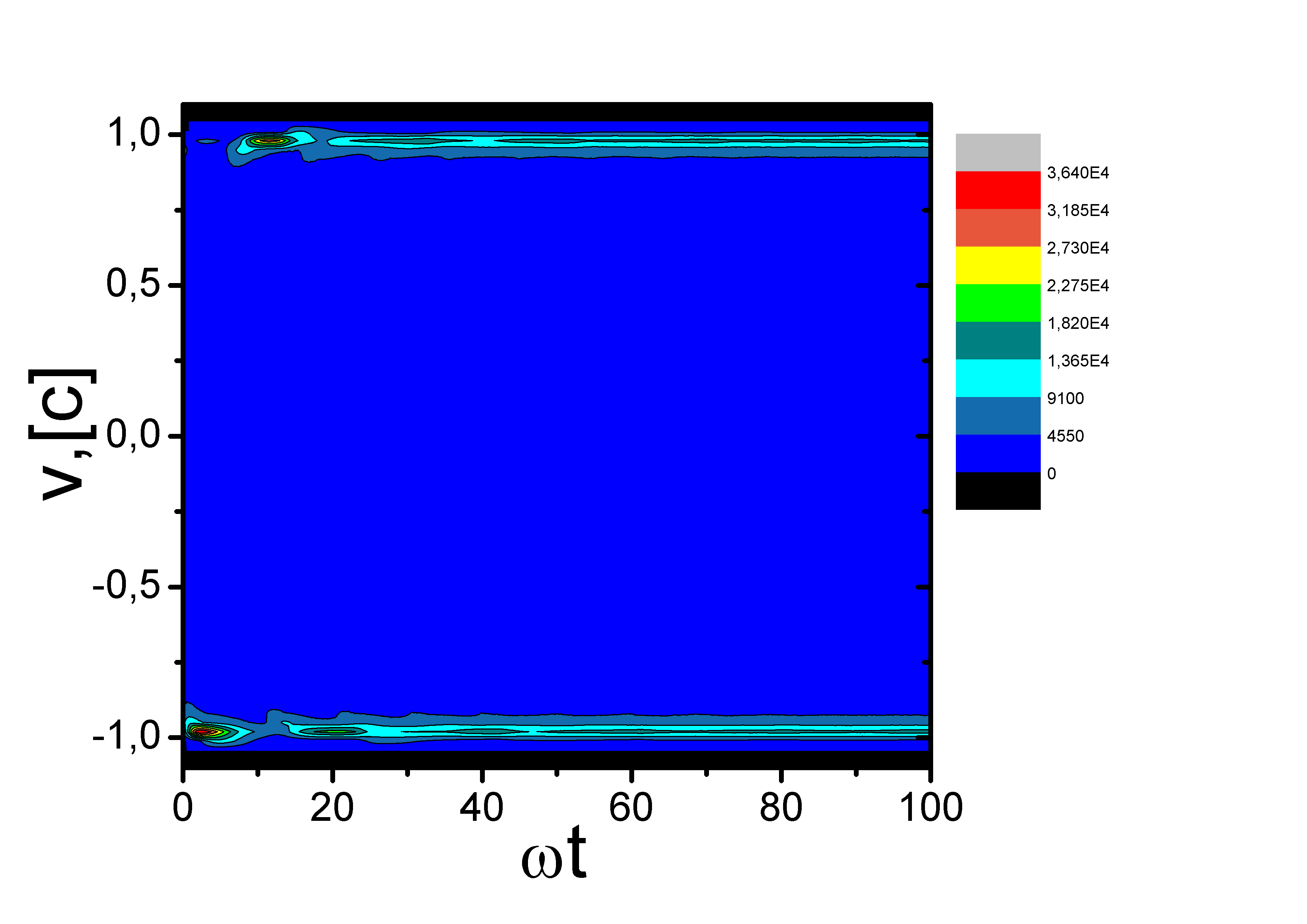}
    \caption{Time evolution of the
non normalized velocity distributions for oscillators at $z=1$ (top) and
$z=0.1$ (bottom). The left top panel -- initial at $t = 0$ (solid line) and
at $\omega t = 86.7$ distributions (dot line). The right
top panel -- the time contour plot, the left bottom panel --
initial at $t = 0$ (solid line) and at $\omega t = 86.7$
distributions (dot line), the right  bottom panel -- the time contour
plot} \label{DstSp}
    \end{figure}

%\subsection{Time evolution of the distribution functions}
{\bf Time evolution of the distribution functions}.
Let us consider four different oscillators with
equal parameter $\omega$, but with different masses $m$, so the
related parameters $z = mc^2/\hbar\omega$ are equal to $100$, $10$,
$1$, $0.1$ (lines $1$, $2$, $3$ and $4$ respectively on figures below).
The time evolution of momentum, coordinate and velocity
distributions are presented
by the %Figs.\ref{DsIm},
Figs.\ref{Imtm},  \ref{Sumq}, \ref{DstSp}. The almost non relativistic
oscillator relates to large value of $z = 100$. As it follows from
Fig.\ref{Imtm} for $z = 100$, the Gaussian shape of momentum
distribution is conserved during the time evolution. The same is
valid for coordinate distribution (not shown). Just on the contrary
the shape of the momentum and coordinate distributions for
relativistic quantum oscillator are considerably changing with time
(Figs.\ref{Imtm},  \ref{Sumq}). Firstly, one can see significant
distribution spreading; secondly, tails of the distributions are
"drawn forward" due to the difference in the  periods of oscillation
of the virtual trajectories with different initial energies
especially for larger values of momentum and coordinate (see
(\ref{TRnut})). Oscillations of the trajectories with larger values
of energy retard by phase from oscillations with lower energies.
This results in appearance of unexpected local maximums ('protuberances')
(Fig.\ref{Imtm}, top left panel).
% thus this "protuberances" form.
Let us stress that the initial $W(p)$ and $W(x)$ ($t = 0$) are the normal Gauss distributions.
% (roughness on the top are statistical error).
%        Now we will consider behavior in time of momentum, velocity and coordinate distributions
%   $W(p)$, $W(v)$ and $W(x)$ . On the fig.~\ref{Imtm} one can see momentum distribution $W(p)$
%    at the initial moment $t = 0$  and at the moment  $t = 86.7/\omega$ for oscillator
%    with $z = 1$ (left figure). This distribution initially is normal
%   (roughness on the top are statistical error). When system evolutes, form of distribution changes:
%    it have many local maximums (fig.~\ref{DsIm}, right panel). Cause of this is considered below.
%   On the Fig.~\ref{Imtm} evolution of $W(p)$ for oscillator $z = 0.1$ is
%    depicted; left panel - distribution W(p) at $t = 0$ and $t = 86.7/\omega$
%    , right panel - contour plot.
%    On the Fig.~\ref{Imtm} Fig.~\ref{Imtm}  one can find momentum and velocity distributions for oscillator
%    with $z = 0.1$ for time moments $t = 0$ and $t = 86.7/\omega$. They behave similar to above;
%    maximum of velocity is nearer to $c$.

Velocity distribution $W(v)$ has an interesting shape presented by the
Fig.~\ref{DstSp} for oscillators with $z = 1$ (top) and $z = 0.1$ (bottom).
Due to the complicated transformation
% \begin{eqnarray}\label{vfromp}
$
        v = \frac{pc}{\sqrt{p^2 + m^2c^2}}
$
positions of the maximum of the velocity
distributions $W(v)$ at initial time $t = 0$ does not coincide with the
position of the maximum of momentum distribution at $p=0$.
Moreover at time evolution the velocity
distribution is restricted by the light speed velocity, while the
momentum distributions has not any limits. Asymmetry of velocity
distribution at the initial moment is the result of errors, which
is related to sampling of the exponentially rare
events in the tails of the momentum Gauss distributions function.
%    It is clear that maximum of the distribution $W(v)$
%    at initial moment $t = 0$ is in point $v = 0.7c$. It can be described by the next way:
%    velocity can't be greater then $c$, although momentum has not any limit. Asymmetry of velocity distribution
%     at the initial moment (and "noise" so) is a result of statistical errors, which was meant above.
%     On the Fig.~\ref{EvdsV} time evolution of W(v) for oscillator
%     with $z = 0.1$ is depicted.
%     On Fig.~\ref{Sumq} one can find evolution of coordinate
%     distribution $W(x)$ for oscillator $z = 1$ (left plot) and $z = 0.1$
%     (right plot).
Take notice that on Fig.~\ref{Imtm} - Fig.~\ref{Sumq} distributions are not normalized on unity.
Its presented values show the number of the virtual trajectories have been counted in the vicinity
%neighborhood
of the each point on the plane related to presented distributions.
% with coordinates on ordinate and abscissa axises plane   of the .

%%%%%%%%%%%%%%%%%%%%%%%%%%%%%%%%%%%%%%%%%%%%
%    \begin{figure}[h]
%    %\vspace{0cm} \hspace{0.0cm}
%%    \includegraphics[width=8.5cm,clip=true]{p_distribution_1.eps}
%    \includegraphics[width=8.5cm,clip=true]{p_distribution_01.eps}
%    \caption{
%    Evolution of momentum distributions in time for oscillators with $z=1$ Х $z=0.1$.
%    }
%\label{Sump}
%    \end{figure}
%    \begin{figure}[h]
%    %\vspace{0cm} \hspace{0.0cm}
%    \includegraphics[width=8.5cm,clip=true]{v_distribution_1.eps}
%    \includegraphics[width=8.5cm,clip=true]{v_distribution_01.eps}
%    \caption{
%           Evolution of velocity distributions in time for oscillators with $z=1$ Х $z=0.1$.
%    }
%\label{Sumv}
%    \end{figure}
%%%%%%%%%%%%%%%%%%%%%%%%%%%%%%%%%%%%%%%%%%%%%

\begin{figure}[htb]
    %\vspace{0cm} \hspace{0.0cm}
    \includegraphics[width=6cm,clip=true]{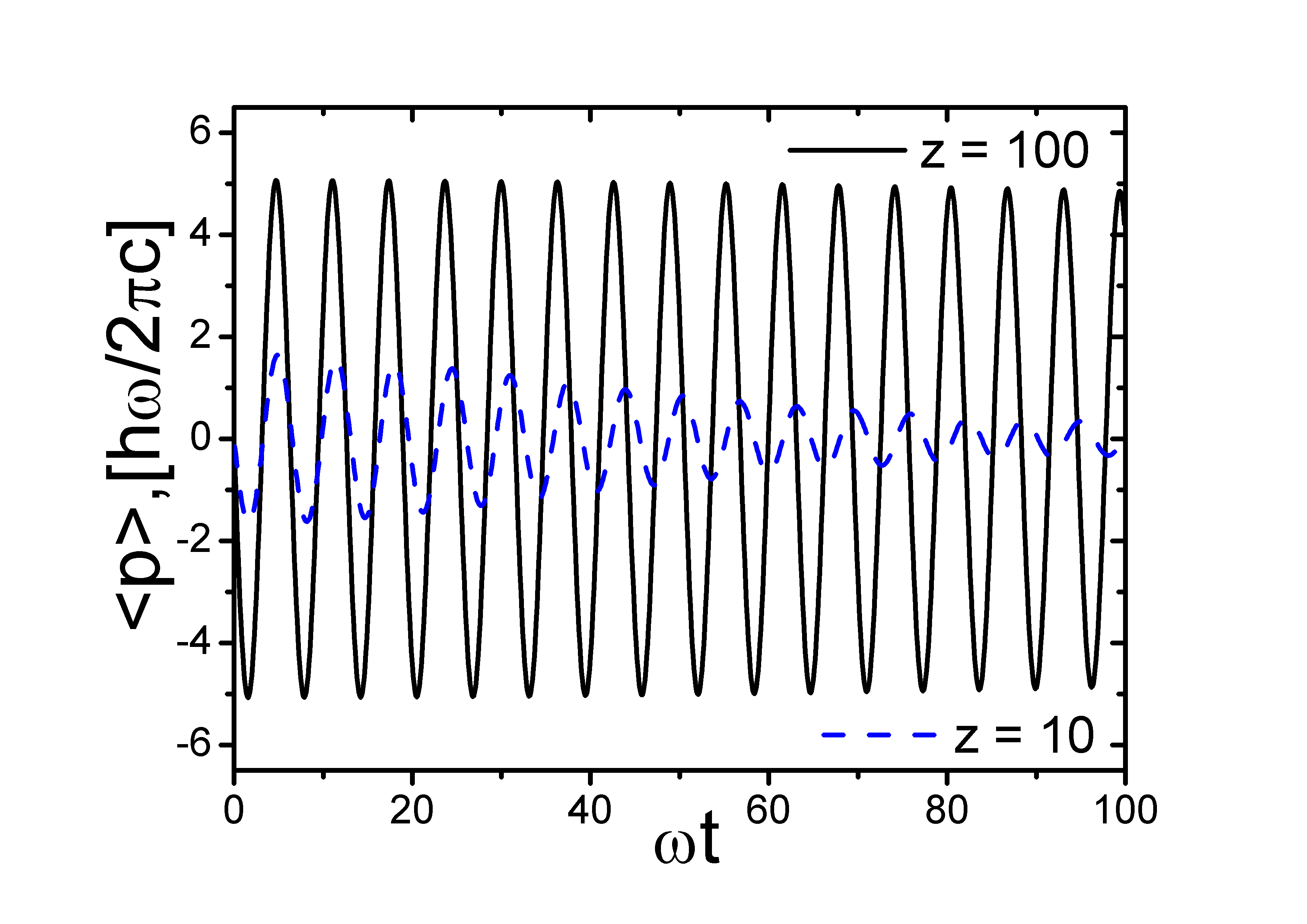}
    \includegraphics[width=6cm,clip=true]{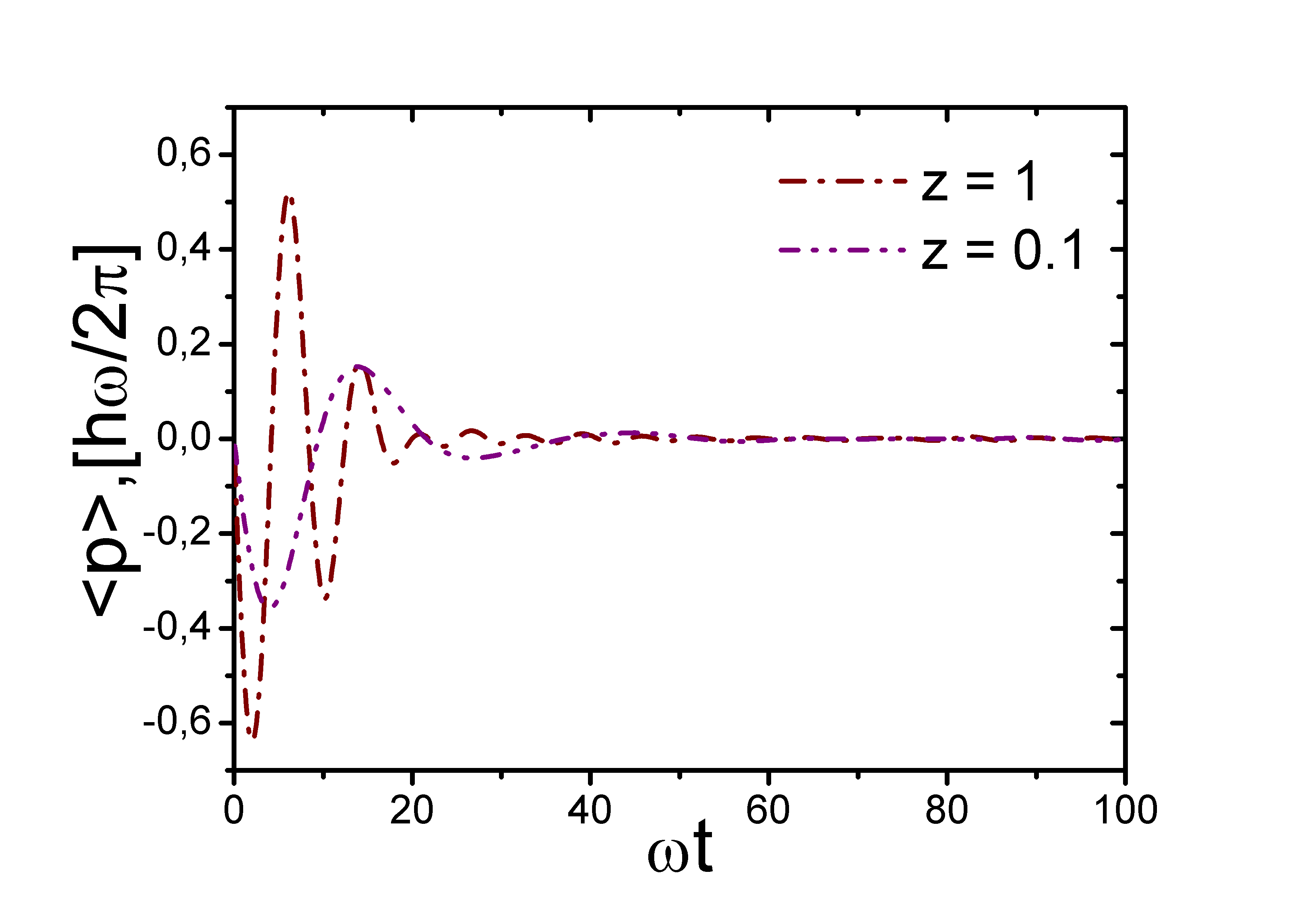}
    \includegraphics[width=6cm,clip=true]{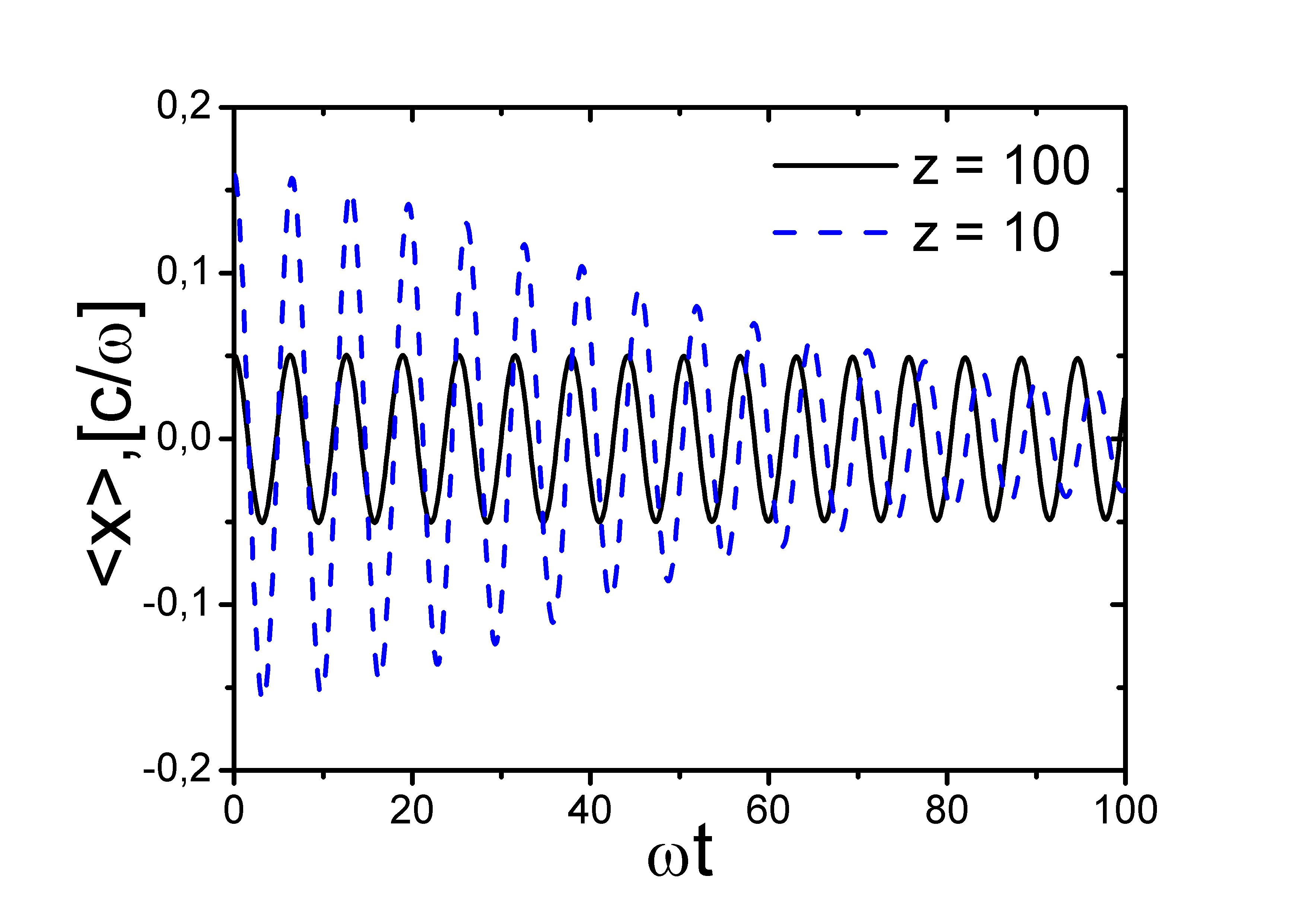}
    \includegraphics[width=6cm,clip=true]{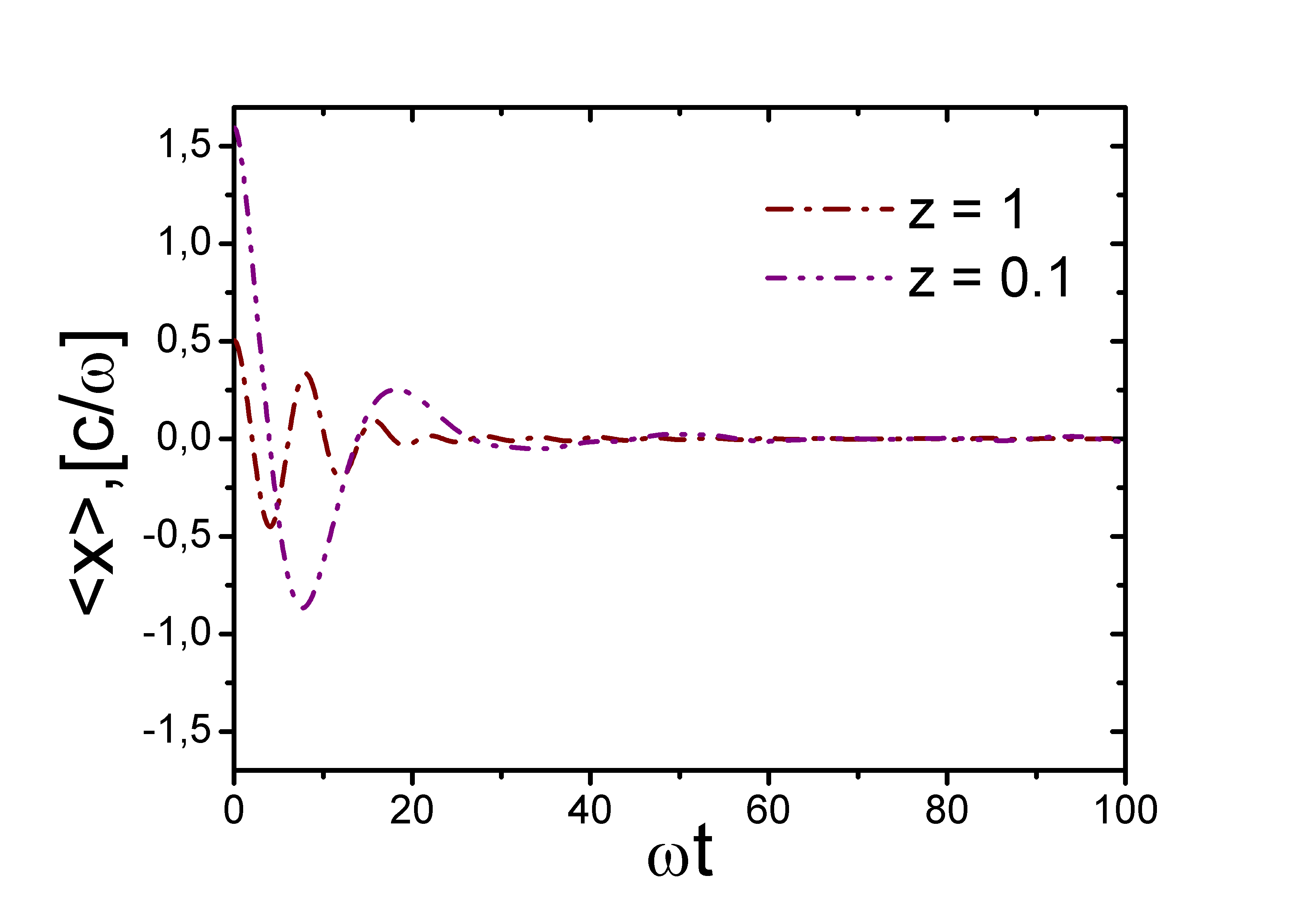}
    \caption{
    Dependence of average momentum $<p(t)>$ and average coordinate $<x(t)>$ on time.
    Lines $1$ correspond to oscillator
    with $z =  100$, line $2$ - $z = 10$, line $3$ - $z = 1$ and line $4$ - $z = 0.1$.
    With decreasing of z period of oscillations rises, while oscillations are damped.
   }
\label{SrImp}
\end{figure}

%\subsection{Average values of quantum operators}
{\bf Average values of quantum operators}.
Now we are going to consider behavior of average values of quantum
operators in time for discussed above initial Wigner function.
%For each oscillator we have used initial Wigner function :
%\begin{eqnarray}\label{IS}
% W_0(x,p) = \frac{1}{\pi\hbar}exp{\left[ -\frac{m\omega(x - \tilde x_0)^2}{\hbar} -
%        \frac{(p - \tilde p_0)^2}{\hbar m\omega} \right]}.
%\end{eqnarray}
On the Fig.~\ref{SrImp} one can see time dependence of average momentum $<p(t)>$ and
the average coordinate $<x(t)>$.
For $z=100$ (weak relativism) $<p(t)>$ and $<x(t)>$ are sinusoidal functions with
period of oscillation equal to $2\pi/\omega$.
Here the average momentum and coordinate behave almost classically like sinusoidal trajectory
 corresponding to initial data $\tilde p(0) =  0$, $\tilde x(0) = \tilde x_0$ \cite{quantum}.
    To analyze the increasing influence of relativistic effects let us consider
     lines $2$,$3$,$4$. Firstly, with decreasing parameter $z$ period of oscillations
is increasing. Secondly, oscillations are damped as the Wigner functions are spreading in
     the phase space.

%     On the Fig.~\ref{SrCr} one can see time behavior of the average coordinate $<x(t)>$.
%   The line $1$ for $<x(t)>$ is sinusoidal function with period $2\pi/\omega$ as for momentum,
%   and the average coordinate behaves almost classically like coordinate for trajectory
%   with initial data $\tilde p(0) =  0$, $\tilde x(t) = \tilde x_0$ .
%    With decreasing of $z$ period of oscillations and decaying are rising. Behavior of $\tilde
%     x(t)$ is like behavior of average momentum (Fig.~\ref{SrImp}); here one can see basic difference
%     from behavior of individual virtual trajectories of semi-relativistic oscillator (in text below).
%\begin{figure}[htb]
%%\vspace{0cm} \hspace{0.0cm}
%\includegraphics[width=6cm,clip=true]{q_1_2.eps}
%\includegraphics[width=6cm,clip=true]{q_3_4.eps}
%\caption{
%      Dependence of average coordinate $\tilde x$ on time. Line $1$ corresponds to oscillator
%    with $z =  100$, line $2$ - $z = 10$, line $3$ - $z = 1$ and line $4$ - $z = 0.1$.
%}
%\label{SrCr}
%\end{figure}

\begin{figure}[htb]
%\vspace{0cm} \hspace{0.0cm}
\includegraphics[width=6cm,clip=true]{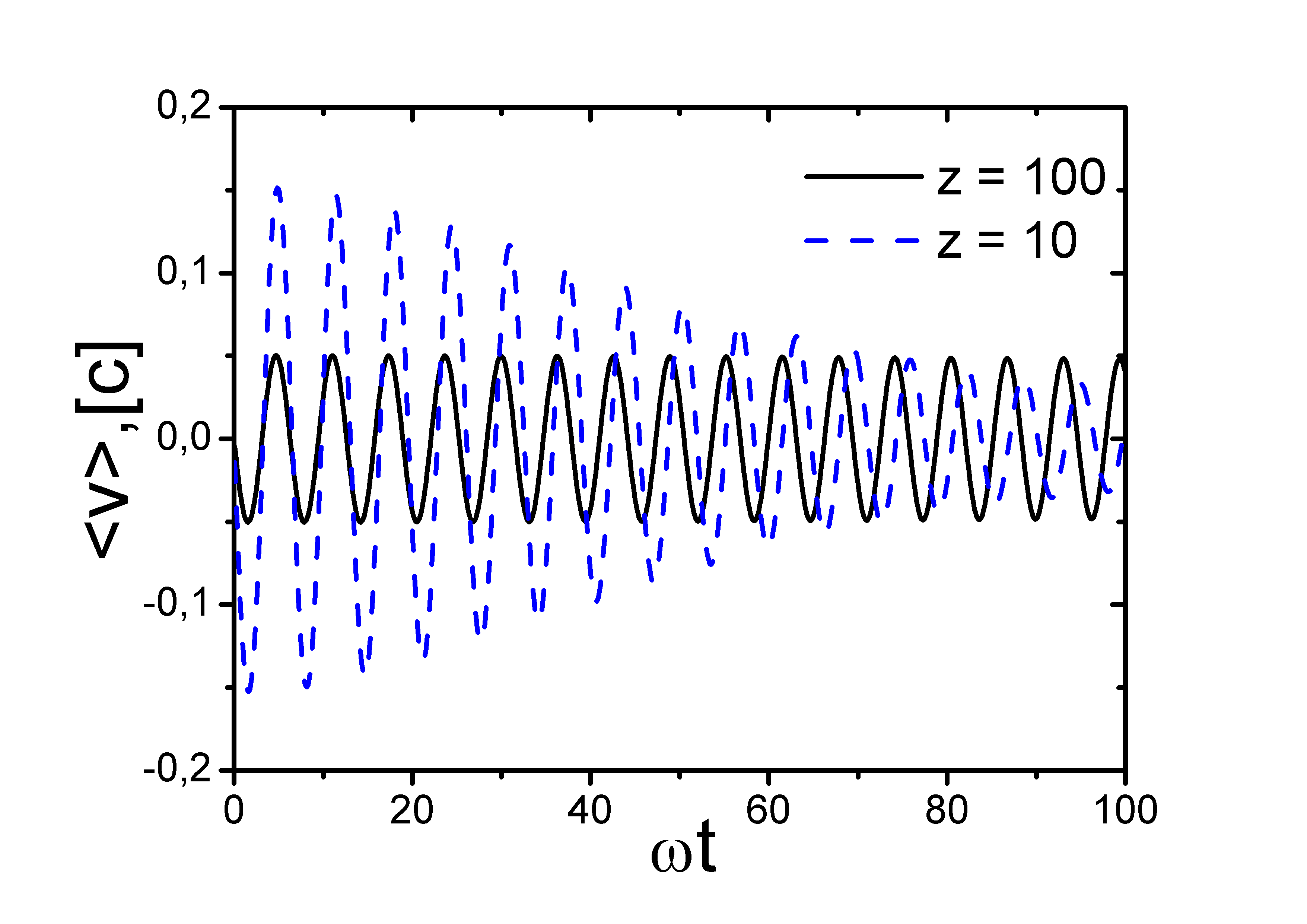}
\includegraphics[width=6cm,clip=true]{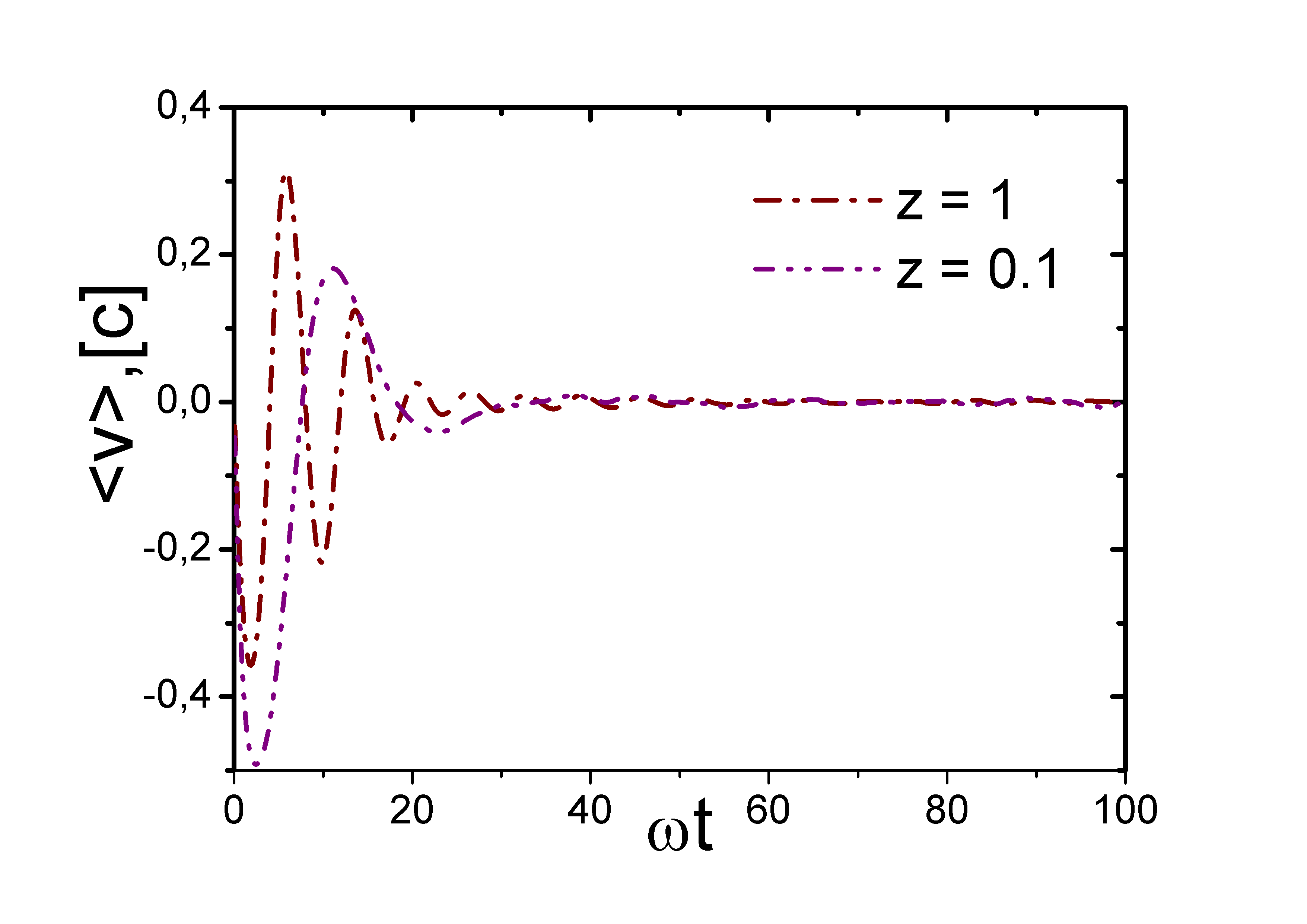}
\caption{
   Dependence of average velocity $<v(t)>$ on time. Line $1$ corresponds to oscillator
    with $z =  100$, line $2$ - $z = 10$, line $3$ - $z = 1$ and line $4$ - $z = 0.1$.
}
\label{SrSp}
\end{figure}

    Fig.~\ref{SrSp} presents the time evolution of the average velocity $<v(t)>$.
    Behavior of the average velocity is similar to behavior of the average momentum and coordinate. Moreover,
    the phases of momentum and velocity oscillations (Fig.~\ref{SrImp}) coincide with each other.
    Let us stress that due to the influence of relativistic effects (lines $2,3,4$)
    $<p>/<v>$ is not equal to mass of oscillator $m$.

\begin{figure}[htb]
%\vspace{0cm} \hspace{0.0cm}
\includegraphics[width=6cm,clip=true]{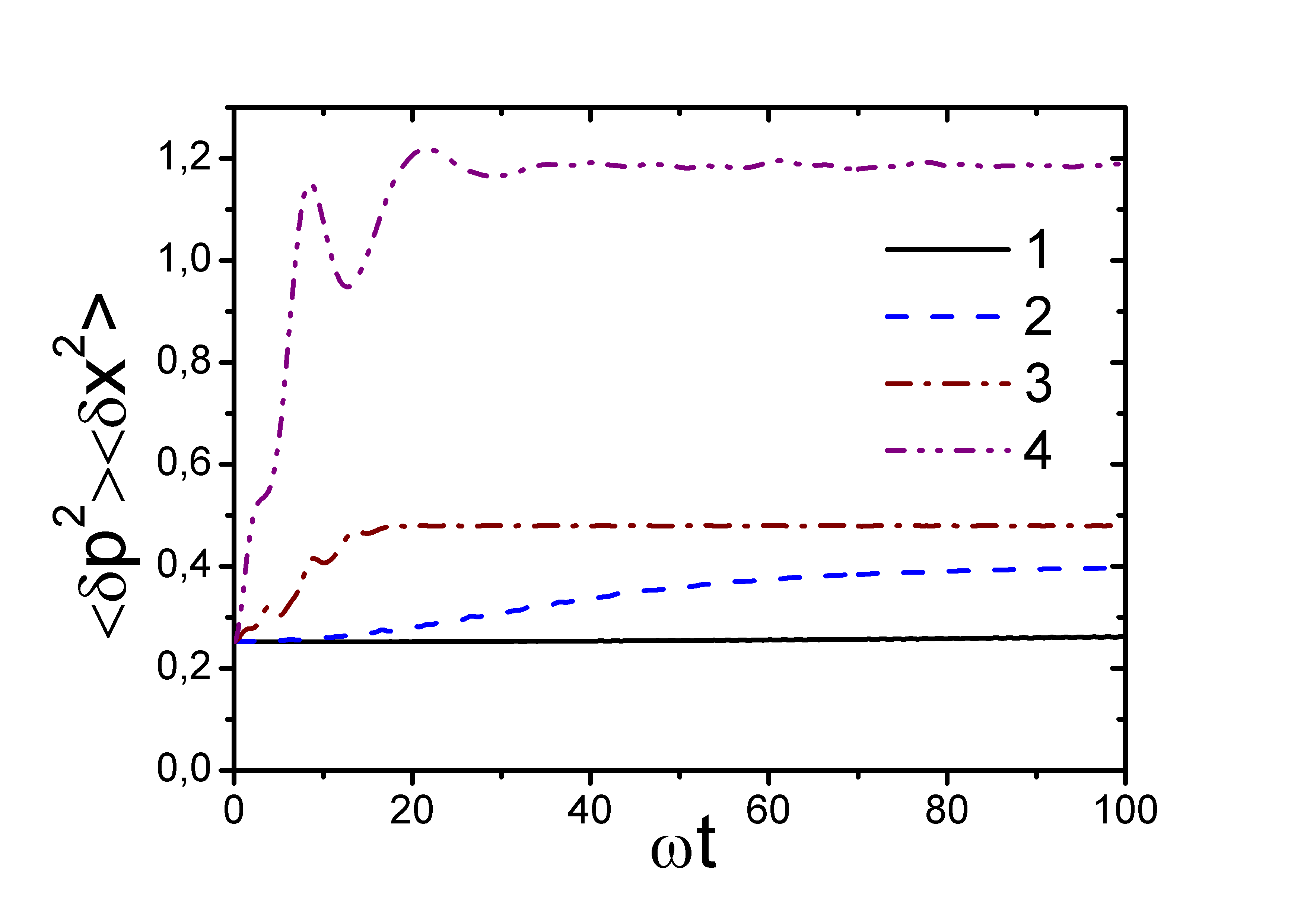}
\includegraphics[width=6cm,clip=true]{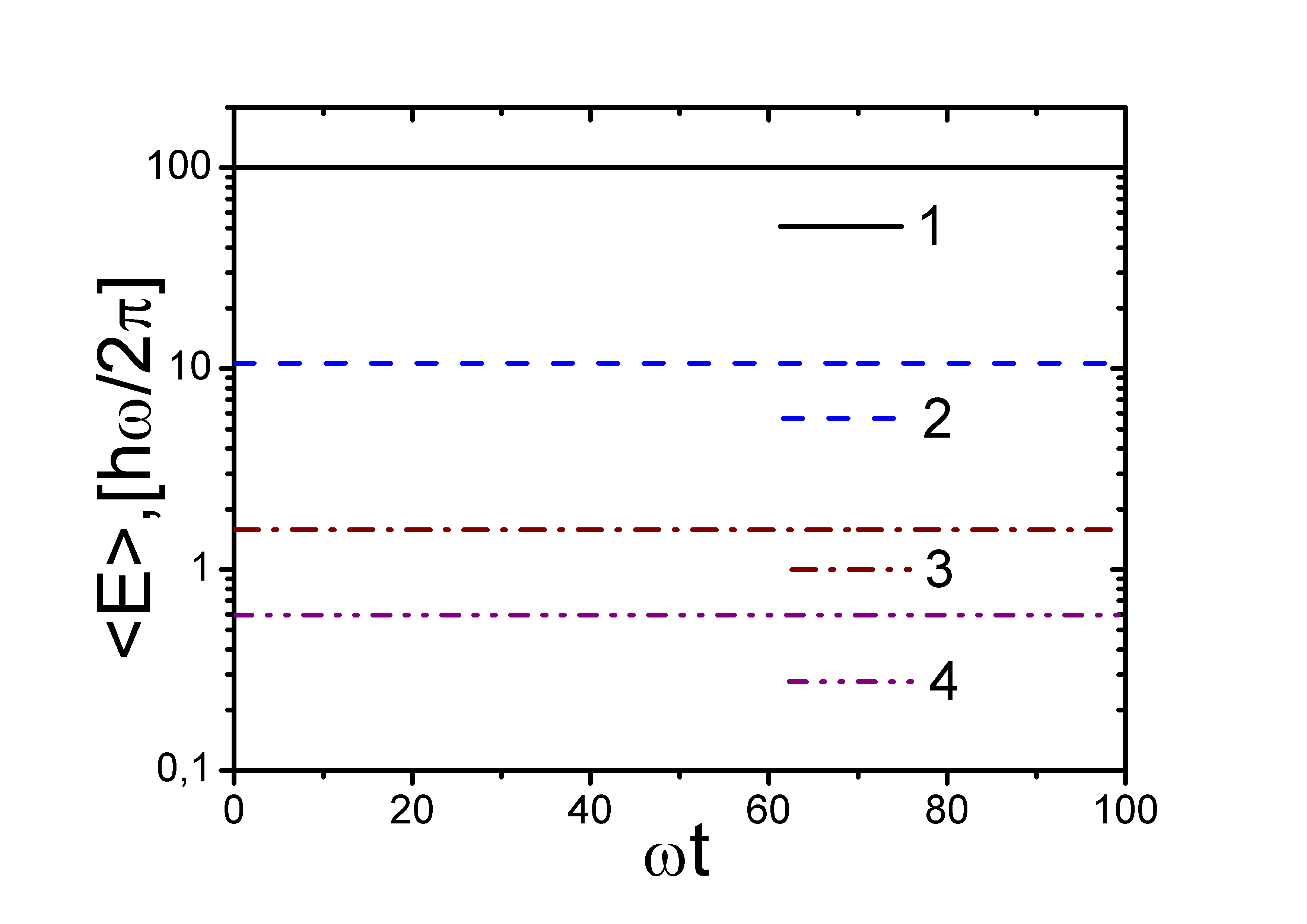}
\caption{Left panel -- dependences on time of the products of the momentum and
coordinate dispersions.
%Line $1$ corresponds to oscillator with $z =  100$, line $2$ - $z = 10$, line $3$ - $z = 1$ and line $4$ - $z = 0.1$.
%    Heisenberg's  indeterminancy principle is satisfied at $t = 0$ it has it's minimum.
Right panel -- dependence of average energy  $<E>$ on time for semi-relativistic oscillators.
%Line $1$ corresponds to oscillator with $z = 100$, line $2$ - $z = 10$, line $3$ - $z = 1$ and line $4$ - $z = 0.1$.
%          It is clear that energies are constants.
}
\label{Hsbr}
\end{figure}

The products of momentum and coordinate dispersions $<\delta p^2><\delta x^2>$
versus time are presented on the left panel of Fig.~\ref{Hsbr}. The Heisenberg's
uncertainty principle is satisfied as
$<\delta p^2><\delta x^2> \ge \hbar^2/4,$. More over these products have
  minimum at $t = 0$ due to the proper choice of initial state.
In case of weak relativism $z =100$ (line $1$) the product of dispersions is
almost constant as the initial Wigner function is not spreading. This results to the
horizontal line presenting the $<\delta p^2><\delta x^2>$.
On the contrary the products of dispersions increase in time very fast
for small $z$. The reason for this is that the initial Wigner function does not
correspond to the coherent or eigen states of the semi-relativistic harmonic
    oscillator.

The right panel of Fig.~\ref{Hsbr} show the dependence of
average energy $<E(t)>$ on time. As the considered oscillators are
conservative the energies are constants.
%Decreasing of energy with rise of $z$ connected with decreasing of rest mass $m$.
For weak relativism (lines $1$, $2$) average energy $<E>$ is circa
the rest energy $mc^2$ ($100\hbar\omega$ and $10\hbar\omega$ ). Due
to the strong relativistic effects (line $4$) the average energy
$<E>$ is considerably greater then rest energy $mc^2$
($mc^2 =0.1\hbar\omega$ for $z=0.1$).

%%%%%%%%%%%%%%%%%%%%%%%%%%%%%%%%%%%%%%%%%%%%%%%%%%%%%%%%%%%%%%%%%%%%%%%%%%%%%%%%%%
%%%%%%%%%%%%%%%%%%%%%%%%%%%%%%%%%%%%%%%%%%%%%%%%%%%%%%%%%%%%%%%%%%%%%%%%%%%%%%%%%%
%%%%%%%%%%%%%%%%%%%%%%%%%%%%%%%%%%%%%%%%%%%%%%%%%%%%%%%%%%%%%%%%%%%%%%%%%%%%%%%%%%
%\begin{figure}[htb]
%    %\vspace{0cm} \hspace{0.0cm}
%    \includegraphics[width=6cm,clip=true]{energy.eps}
%    \caption{
%         Dependence of average energy  $E$ on time for semi-relativistic oscillators.
%         Line $1$ corresponds to oscillator with $z =
%           100$, line $2$ - $z = 10$, line $3$ - $z = 1$ and line $4$ - $z = 0.1$.
%          It is clear that energies are constants.
%    }
%\label{SrEn}
%\end{figure}

%%%%%%%%%%\newpage

%\subsection{Virtual trajectories}
{\bf Virtual trajectories}.
To understand physical reasons of the different properties of
non-relativistic and relativistic harmonic oscillators we have to
consider behavior of individual virtual trajectories related to
initial delta distribution with fixed initial momentum and
coordinate
\begin{eqnarray}\label{Delta}
        W_0(x_0,p_0;0) = \delta(x -x_0)\delta(p - p_0),
\end{eqnarray}
where $(p_0, x_0)$ are
%$ \bar x(t)$,$ \bar p(t)$ - are the  with
initial condition of the virtual trajectories. Here due to the limitation (\ref{ro})
distribution $W_0$ can not be considered as physical Wigner function.
However this choice of the Wigner function can help in understanding
peculiarities in the time behavior of the discussed above average
values.
%To analyze the influence of relativistic effects we have considered
%two oscillators with equal $\omega$ but different masses related to the different
%$z$.
The given choice of the Wigner function (\ref{Delta})
reduces the averaging over ensemble of the virtual trajectories
to the consideration of the contribution only one virtual trajectory with fixed initial
$x_0,p_0=0$ like in classical case.

\begin{figure}[htb]
    %\vspace{0cm} \hspace{0.0cm}
    \includegraphics[width=5.91cm,clip=true]{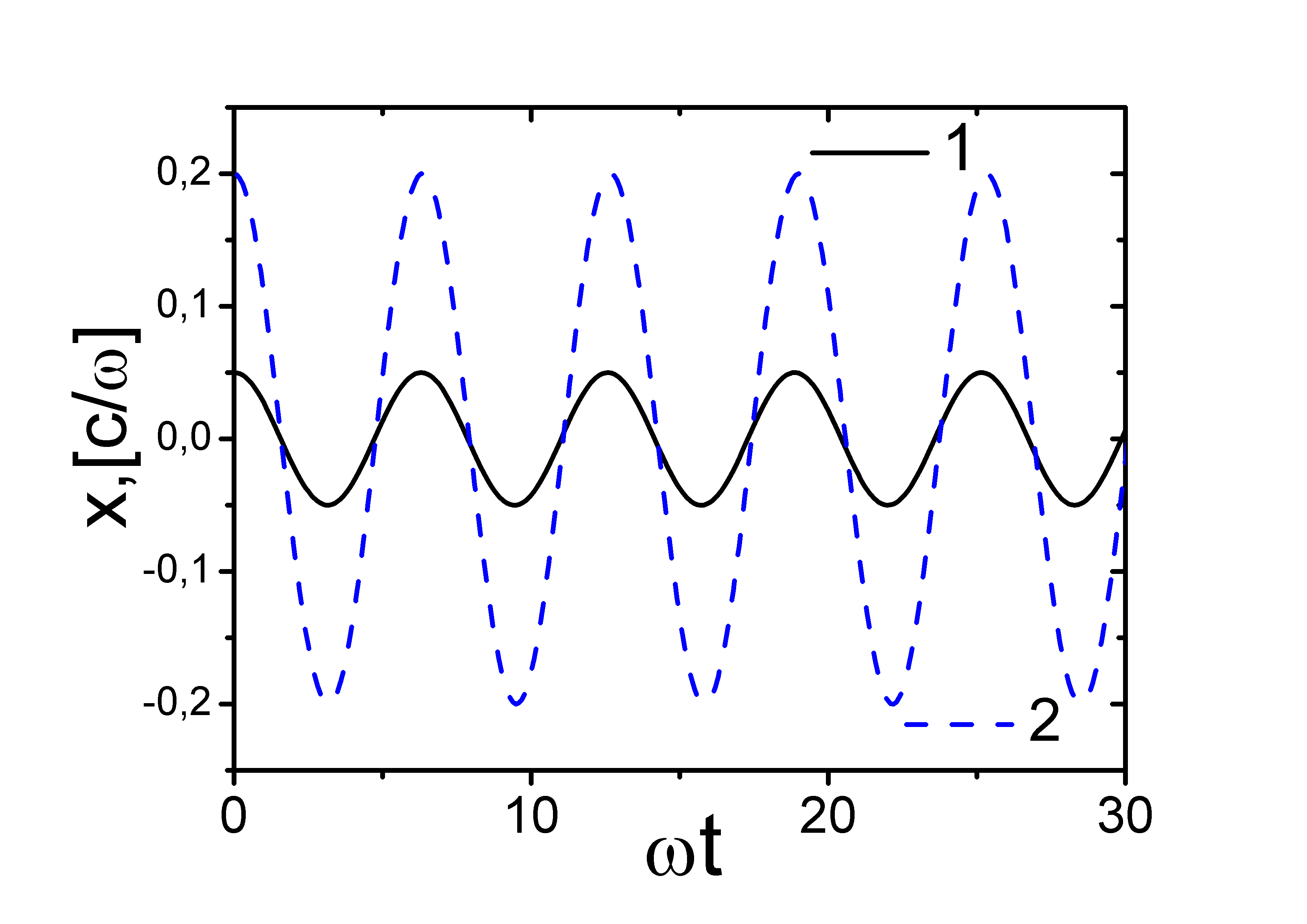}
    \includegraphics[width=5.91cm,clip=true]{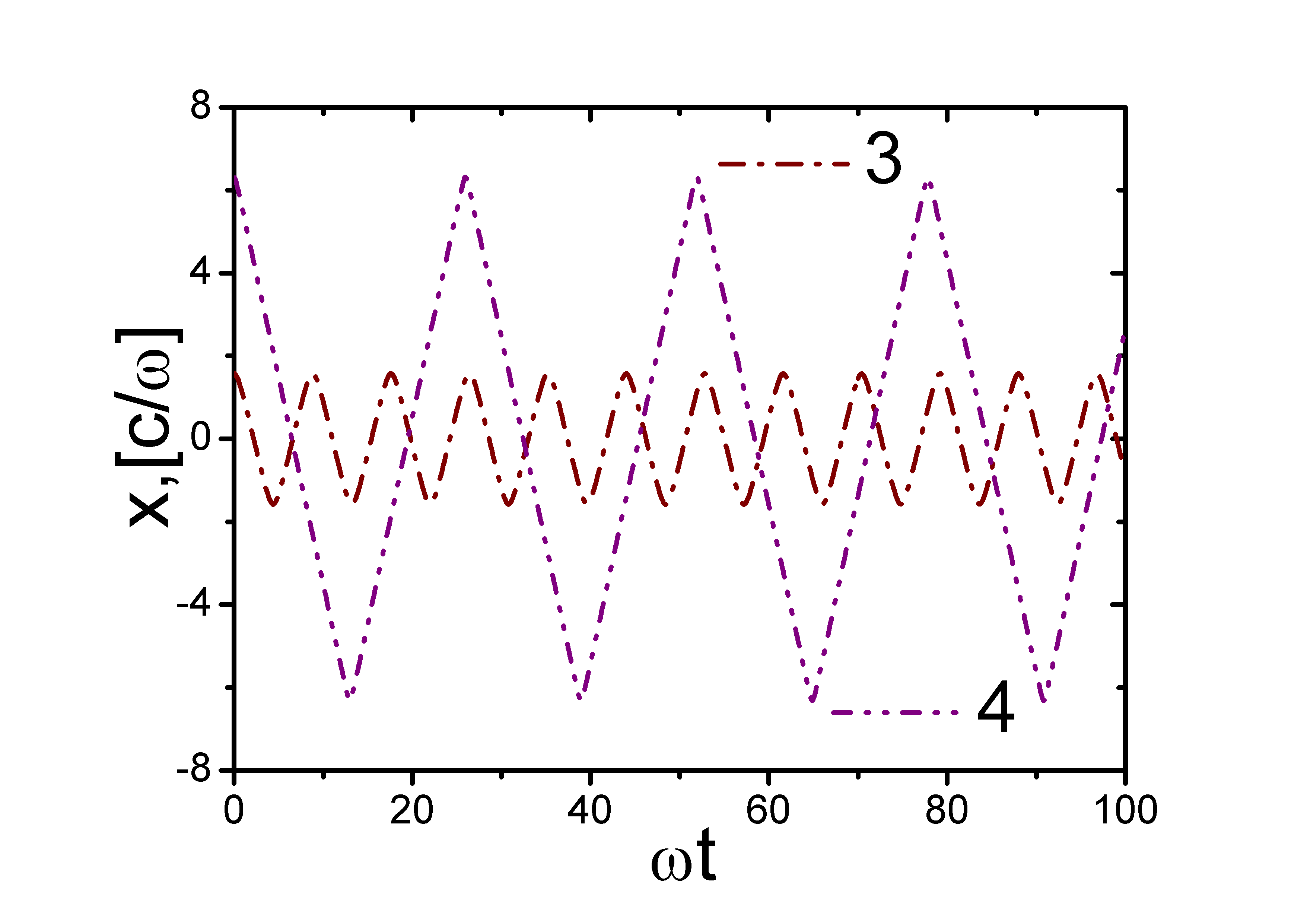}
    \includegraphics[width=5.91cm,clip=true]{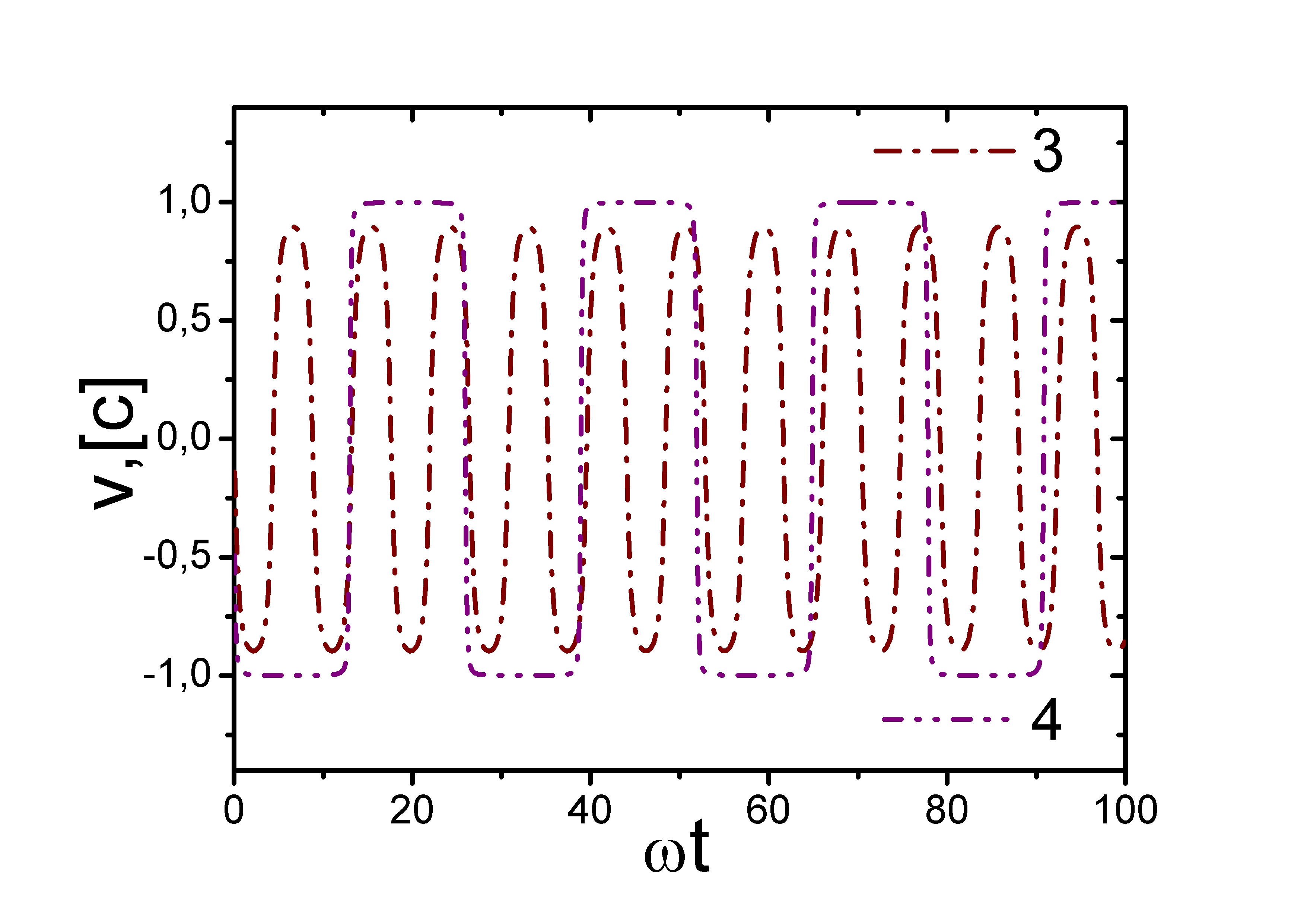}
    \caption{
Virtual trajectories for $p_0=0$. Left panel -  non-relativistic trajectories $1$
(with $x_0=0.05 c/\omega $) and $2$ (with $x_0= 0.2 c/\omega $) for $z=100$.
Central and right panels - relativistic trajectories $3$ (with $x_0=1.58 c/\omega$)
and $4$ (with $x_0=6.3 c/\omega$) $z =0.1$.
%Left panel - $\bar{x}(\omega t)$ for $z=100$ and $p_0=0$ but with different
%$x_0=0.2, 0.05$ (lines $1$ and $2$ respectively). Central and right panels -
%for $z = 0.1$ $\bar{x}(\omega t)$ and $\bar{v}(t)$ respectively; lines $3$ and $4$
%for $p_0=0$ but with different $x_0=0.2, 0.05$,
%Right panel -- velocity $\bar{v}(t)$.
% of the virtual trajectories $3$, $4$ ($z = 0.1$).
}
\label{cr}
\end{figure}

Fig.~\ref{cr} presents the virtual trajectories as function of time.
For $z = 100$ non-relativistic trajectories look like sinusoid curves
with equal oscillation periods (trajectories $1$, $2$ on the left panel).
Let us note that energy of the virtual trajectory is proportional to the
square of $x_0$ and in the relativistic case (central and right panels $z = 0.1$)
oscillation period strongly increases with growth of the energy of trajectory
(see (\ref{TRnut})).
%For non-relativistic trajectories $1$ (with $x_0=
%0.05 c/\omega $) and $2$ (with $x_0= 0.2 c/\omega $) parameter $z$
%was equal to $100$, while for relativistic ones $3,4$ ($x_0$ equals
%to $1.58 c/\omega$ and $6.3 c/\omega$) parameter $z$ was small ($z = 0.1$)
Period of oscillations of $x(\omega t)$ for curve $4$ is longer than
for one $3$ due to the larger value of energy.
Moreover, curves $3$ and $4$ tend to become the zig - zag lines.

Velocity of the trajectory $4$ (right panel of the Fig.~\ref{cr}) looks
like rectangular wave for high energy. This figure
%Fig.~\ref{cr} (right panel)
illustrates that mostly the virtual trajectory has velocity $v$ approaching
the velocity of
light $c$ and only near the turning points $v$ differs from this
limit.
%Interesting is that velocity of the relativistic trajectory $4$
%tends to rectangular wave with the horizontal parts
%equal almost to $c$.
Velocity of trajectory $3$ for the same $z$ but lower energy
differs considerably from rectangular wave.
%\begin{figure}[htb]
%    %\vspace{0cm} \hspace{0.0cm}
%%    \includegraphics[width=8.5cm,clip=true]{v_tr_100.eps}
%    \includegraphics[width=6.0cm,clip=true]{v_tr_01.eps}
%    \caption{
%Velocity $\bar{v}(t)$ of the virtual trajectories $3$, $4$
%($z = 0.1$). Velocity of the trajectory $4$ with higher energy looks like rectangular wave.
%}
%\label{spd}
%\end{figure}

\begin{figure}[htb]
    %\vspace{0cm} \hspace{0.0cm}
    \includegraphics[width=5.5cm,clip=true]{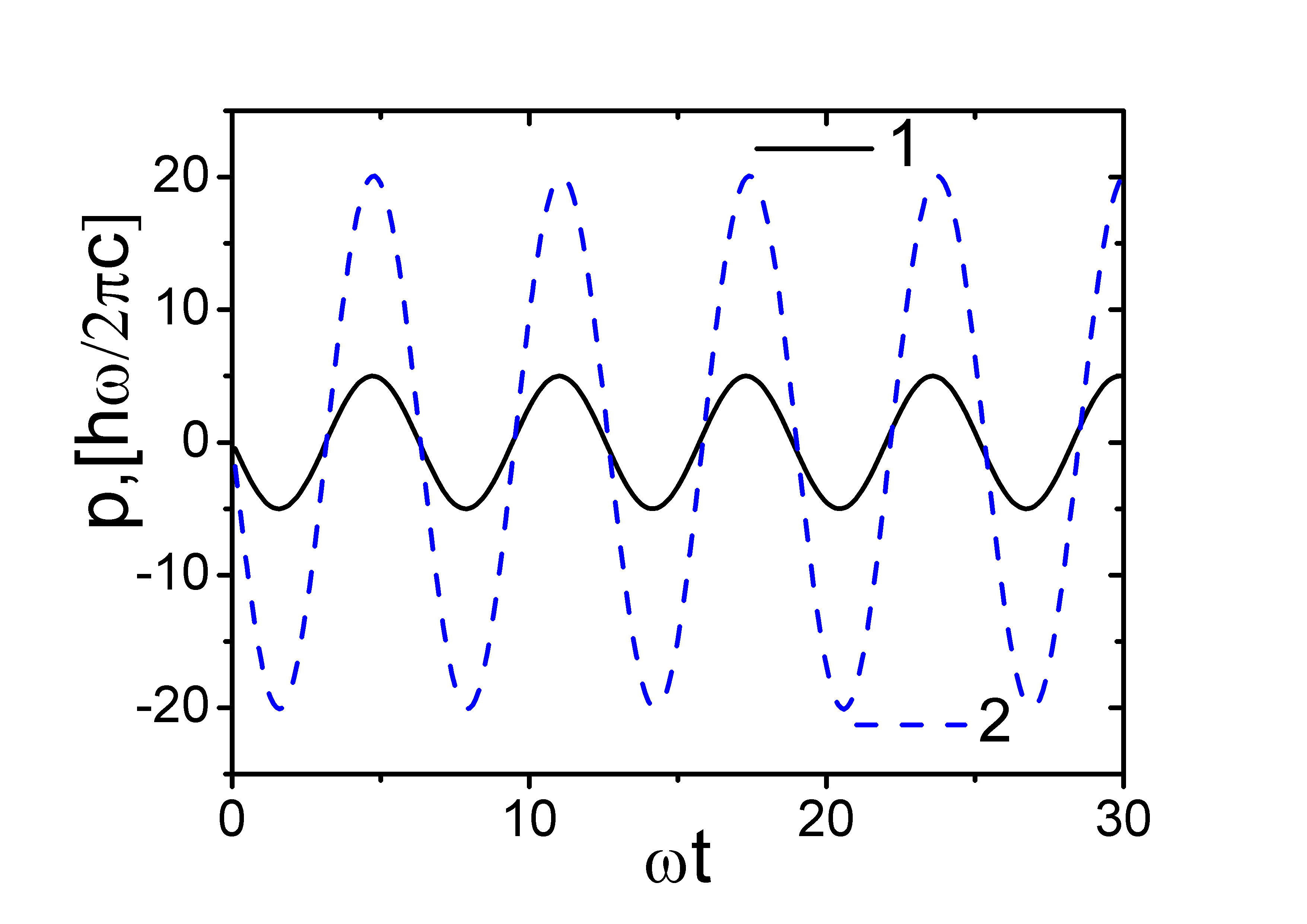}
    \includegraphics[width=5.5cm,clip=true]{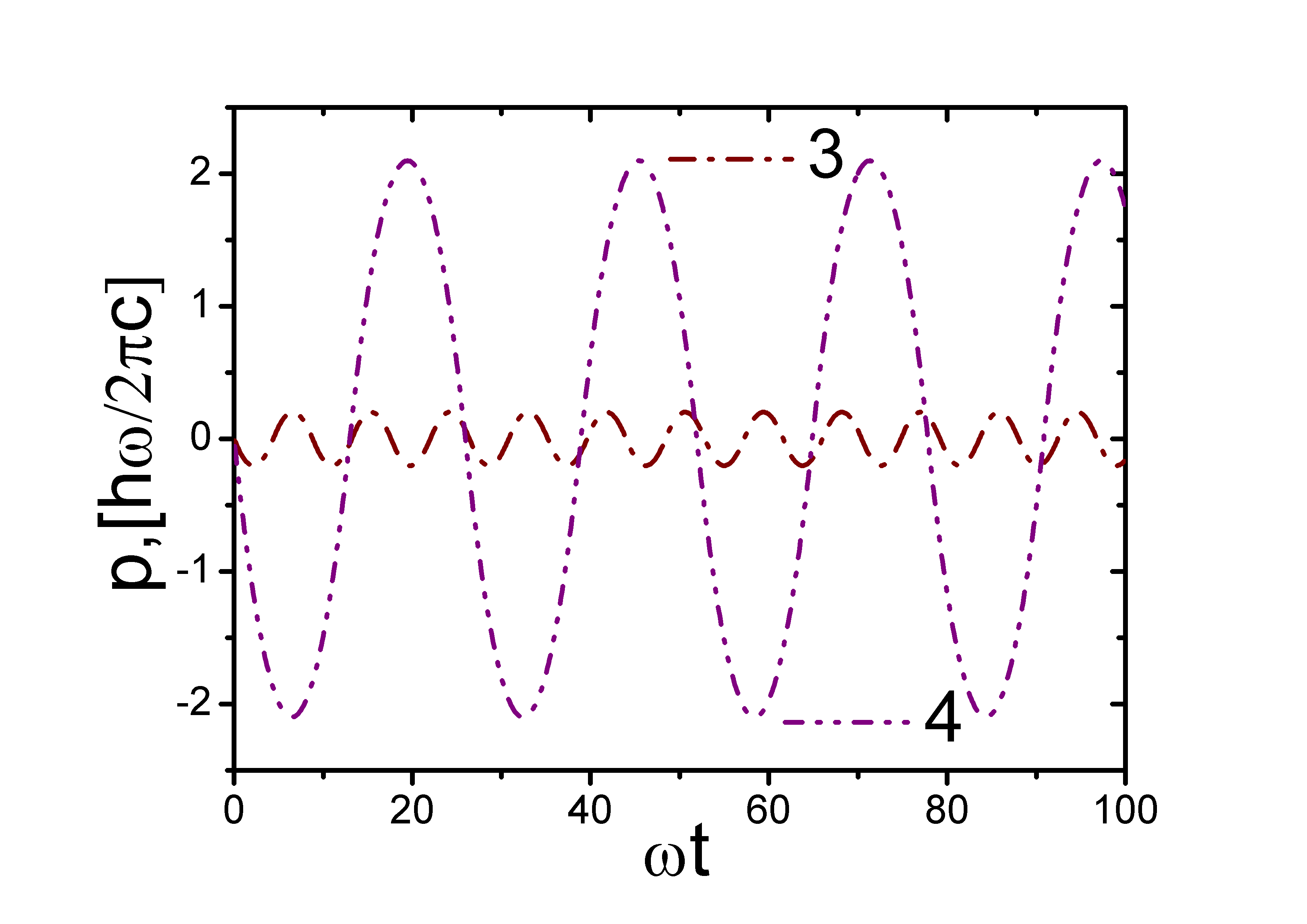}
    \includegraphics[width=5.5cm,clip=true]{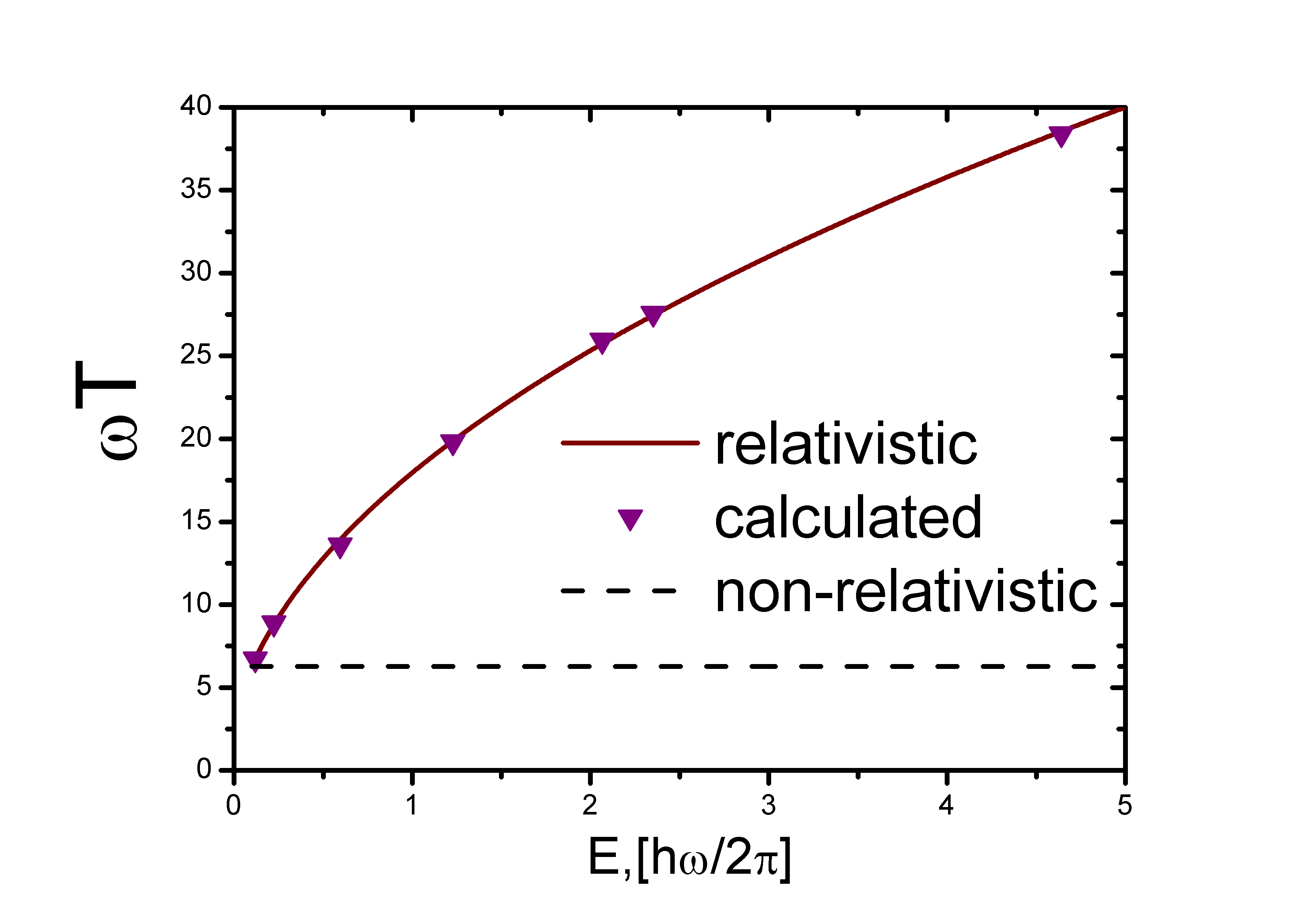}
    \caption{
Virtual trajectories for $p_0=0$. Left panel - non relativistic trajectories $1$, $2$,
central panel -  relativistic ones $3$ and $4$. Oscillation period of relativistic trajectory $4$ is
longer than $3$ due to the larger value of virtual energy.
Right panel - period of oscillations of the virtual trajectories versus energy for $z=0.1$.
Solid line - formula (\ref{TRnut}), triangles - results of numerical simulations,
dashed line - non-relativistic case.
}
\label{imp}
\end{figure}

Left and cetral panels of the Fig.~\ref{imp} show non-relativistic momentum virtual trajectory
 versus time for $z = 100$ (lines $1,2$) and relativistic trajectories for $z = 0.1$
 (lines $3,4$). The time dependences of $p(\omega t)$
look practically as sinusoid curves. In the left panel  period of momentum
oscillations for trajectories $1$ and $2$ is almost the same.
%(period of momentum oscillations of $2$ is a bit longer due to the very small relativistic effect.
In the central panel oscillation period is longer for the relativistic
trajectory $4$ with higher energy.
% (quantitative comparison will be given  below).
%These results agree with formula (\
%ref{TRnut})(quantitative comparison will be given  below).
%

%On the fig.~\ref{time} one can find relationship of proper time (in
%noninertial frame of reference, associated with relativistic
% particle in oscillator potential) and laboratory frame of reference
% (inertial) in case of oscillator with $z = 0.1$. Here are five
% trajectories, differing from each other by initial energy: $a$ is
% trajectory for resting particle (proper time equals lab time), $b$ is
% trajectory with energy $E_{tr} = 0.13125\hbar\omega$, $c$ is
% trajectory with energy $E_{tr} = 0.225\hbar\omega$, $d$ is
% trajectory with energy $E_{tr} = 0.6\hbar\omega$,$e$ is
% trajectory with energy $E_{tr} = 2.1\hbar\omega$. The figure represents time
% dilation for oscillator.
%\begin{figure}[htb]
    %\vspace{0cm} \hspace{0.0cm}
%    \includegraphics[width=8cm,clip=true]{time_eig.eps}
%    \caption{ Time dilation ( relativistic oscillator $z = 0.1$) for
%    trajectories with different energies.
%.} \label{time}
%\end{figure}

Comparison of analytical and numerical calculations of the oscillation period
is presented by the right panel of the Fig.~\ref{imp}. Analytical dependence
of oscillation period on energy according to the formula (\ref{TRnut})
 is plotted by solid curve for $z = 0.1$. Results of numerical
calculations presented by triangles agree very well with analytical dependence.
%\begin{figure}[htb]
%    %\vspace{0cm} \hspace{0.0cm}
%    \includegraphics[width=8cm,clip=true]{dispersion.eps}
%    \caption{
%Period of oscillations of the virtual trajectories versus energy for $z=0.1$.
%Solid line - formula (\ref{TRnut}), triangles - results of numerical simulations,
%dashed line - non-relativistic oscillator.
%}
%\label{disp}
%\end{figure}
%Testing the energy conservation low for non-relativistic and
%relativistic oscillators are presented by Fig.~\ref{En}. As the
%system is conservative the energies of the virtual trajectories
%versus time are practically constant both for $z = 100$ ($1$ and
%$2$) and for $z = 0.1$ ($3$ and $4$). Energies of the trajectories
%$1$ and $2$ with weak relativism ($z=100$) are almost equal to the
%rest energy $mc^2$ (in this case $100\hbar\omega$ ), while energies
%of the relativistic trajectories $3$ and $4$ ($z=0.1$) differs
%considerably from each other.

%\newpage

%\begin{figure}[h]
%\vspace{0cm} \hspace{0.0cm}
%    \includegraphics[width=7cm,clip=true]{energy_tr_100.eps}
%    \includegraphics[width=7cm,clip=true]{energy_tr_01.eps}
%    \caption{
%Energies of the virtual trajectories versus time for non-relativistic
%($z = 100$, left panel) and relativistic ($z = 0.1$, right panel) oscillators.
%}
%\label{En}
%\end{figure}

\begin{figure}[htb]
    %\vspace{0cm} \hspace{0.0cm}
    \includegraphics[width=6cm,clip=true]{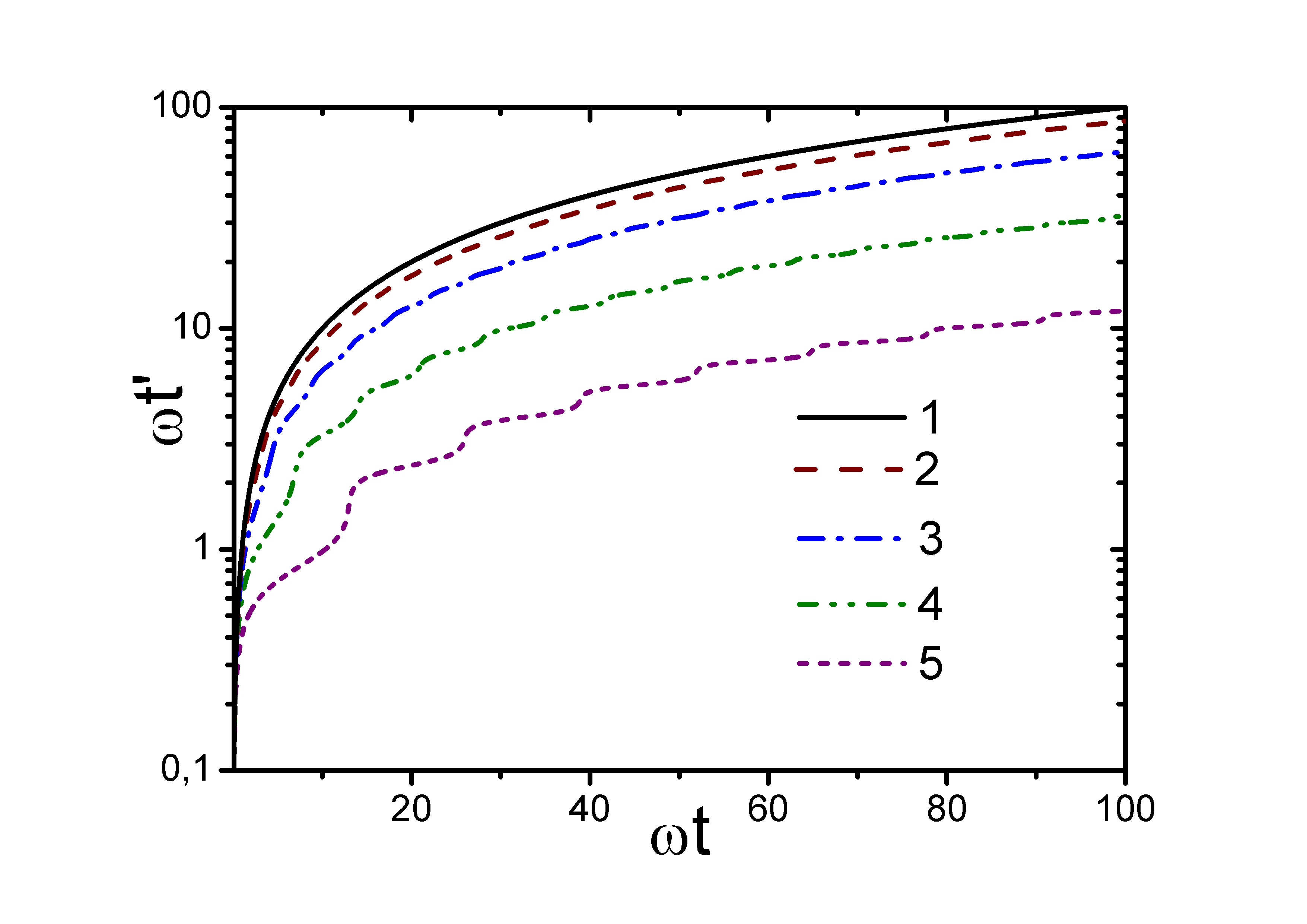}
    \includegraphics[width=6cm,clip=true]{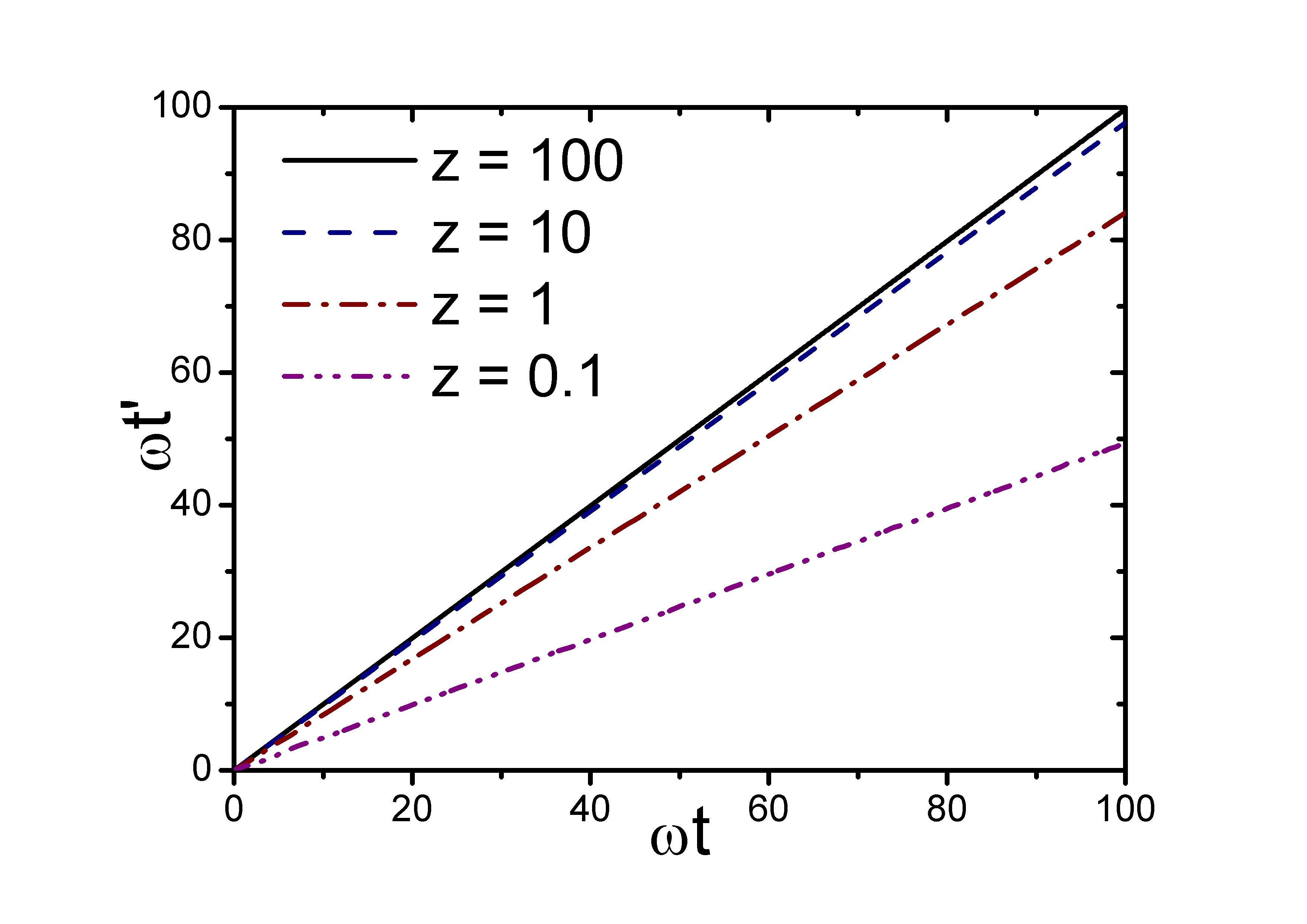}
    \caption{ Time dilation for $t_0=0$.
Left panel - the time dilation
%in logarithmic scale
for virtual trajectories with $z =0.1$.
%(oscillator $z =0.1$) is depicted.
Energies $E_{tr}$ of the virtual trajectories: $1$ -
$0.1\hbar\omega$ (resting particle); $2$ - $0.13125\hbar\omega$; $3$
- $0.225\hbar\omega$; $4$ - $0.6\hbar\omega$; $5$ -
     $2.1\hbar\omega$. Right panel - the time dilation for quantum
    semi-relativistic oscillators. % with different $z$ is depicted.
} \label{time}
\end{figure}

%%\newpage
%\newpage
%\subsection{Time dilation}
{\bf Time dilation}.
One of the most important relativistic effects is the time dilation:
 proper time of relativistic particle is slower than time in the non -relativistic
 lab frame of reference (inertial). Now we will consider this effect for
 semi-relativistic harmonic oscillator.
We begin our consideration for one virtual trajectory
$p(t),x(t)$.The relationship between proper time and lab time is
well known for the case, when particle moves with constant velocity:
\begin{eqnarray} \label{UTD}
           t' - t'_0 = (t - t_0)\sqrt{1 - \frac{\dot x(t)^2}{c^2}},
\end{eqnarray}
where $\dot x(t)$ is velocity of the particle, $t - t_0$ is time interval
between two events in lab frame
of reference, $ t' - t'_0$ is time interval between this two
events in the rest frame of particle. In case of oscillator, however, particle's velocity is not a
constant and formula (\ref{UTD}) is right only for infinitesimal time
intervals:
\begin{eqnarray} \label{ITD}
           dt' = dt\sqrt{1 - \frac{\dot x(t)^2}{c^2}}
\quad \mbox{while for finite time interval } \quad
t' - t'_0 = \int\limits_{t_0}^{t}{dt}{\sqrt{1 - \frac{\dot
    x(t)^2}{c^2}}}.
\end{eqnarray}
%Thus, for finite time interval $t - t_0$ we have next formula:
%\begin{eqnarray} \label{NUTD}
%    t' - t'_0 = \int\limits_{t_0}^{t}{dt}{\sqrt{1 - \frac{\dot
%    x(t)^2}{c^2}}}.
%\end{eqnarray}
Dependence on time for virtual trajectories with different energies
is represented in logarithmic scale by the left panel of Fig.~\ref{time}
for $z = 0.1$. Almost constant parts of the curves ((4),(5)) related to the trajectories
with high energy and correspond to the motion with velocity of order of the
speed of light $c$.

In case of semi-relativistic oscillator with Hamiltonian
(\ref{ORHAM}) one has to average integral in (\ref{ITD}) over all virtual
trajectories. So for quantum oscillator described by the  Wigner function $W(x,p;t)$
% According to formulas (\ref{TDO}),(\ref{weyl}) and (\ref{SrA}),
the time dilation for semi-relativistic oscillator can be calculated by formula:
\begin{eqnarray} \label{QTD}
    t' - t'_0 = \int\limits_{t_0}^{t}{dt}\int{W(x,p;t)\frac{1}{\sqrt{1 + (p/mc)^2}}}{dp}{dx}
\quad \mbox{related to the 'time dilation operator'- } \quad
d\hat t' = dt\sqrt{1 - \frac{\hat{\dot x}(t)^2}{c^2}}.
\end{eqnarray}
%
%Therefore the time dilation can be described by "time dilation operator"
%\begin{eqnarray} \label{TDO}
%           d\hat t' = dt\sqrt{1 - \frac{\hat{\dot x}(t)^2}{c^2}}.
%\end{eqnarray}
In numerical calculation
%we have used a sum instead of integral on time;
the averaging has been done over %by direct summing of $1/\sqrt{1+(p/mc)^2}$ on
all virtual trajectories for each time moment $t$.
Results are represented by the right panel of the Fig.~\ref{time}
for oscillators with $z = 100$, $10$,$1$ and $0.1$.

%%%%%%%%%%%%%%
%----------------------------

\newpage

\section{Conclusion}\label{s:discussion}

In this paper we are doing the exact simulation of time evolution of
semi-relativistic quantum 1D harmonic oscillator. To solve the
Wigner - Liouwille equation for such system we combine Monte-Carlo procedure
and molecular dynamics methods. As initial Wigner quasi distribution function
we have used a coherent state of appropriate non relativistic harmonic oscillator.
We have studied the time evolution of the momentum, velocity and
coordinate Wigner distributions and average values of quantum operators.
Obtained results demonstrates, that relativistic treatment results in
the appearance of the new physical effects as opposed to non-relativistic
case. Interesting is the complete changing of the shape of the momentum,
velocity and coordinate distribution functions as well as formation of
"unexpected protuberances".
% on the contour plot.
We have also calculated relativistic time dilation for oscillator, as it can be useful
for consideration of the life -time of the particle bound states in the traps.

%----------------------------
\bibliographystyle {apsrev}

\end{document}